\documentclass[aps,prd,twocolumn,prd,showpacs,showkeys,preprintnumbers,bibnotes,floatfix,longbibliography,notitlepage,nofootinbib,superscriptaddress,superscriptgaddress,amsmath
]{revtex4-2}

\usepackage{graphicx}
\usepackage{dcolumn}
\usepackage{bm}
\usepackage{tabularx}

\usepackage[colorlinks]{hyperref}
\hypersetup{
    pdfnewwindow=true,      
    colorlinks=true,       
    linkcolor=blue,          
    citecolor=blue,        
    filecolor=blue,      
    urlcolor=blue        
}

\usepackage{booktabs}
\usepackage{xcolor}
\usepackage{listings}
\usepackage{calligra}
\usepackage{xspace}
\usepackage{multirow}
\usepackage{float}

\DeclareMathAlphabet{\mathcalligra}{T1}{calligra}{m}{n}
\DeclareFontShape{T1}{calligra}{m}{n}{<->s*[2.2]callig15}{}

\newcommand{\sun}{ M_\odot }

\begin{document}


\title{A full chemical network treatment of supermassive black hole seed formation from dark matter annihilation or decay}

\author{Han Wu}
\email{han.wu@queensu.ca}
\affiliation{
Department of Physics, Engineering Physics and Astronomy,\\
Queen’s University, Kingston ON K7L 3N6, Canada
}
\affiliation{%
Arthur B. McDonald Canadian Astroparticle Physics Research Institute, Kingston ON K7L 3N6, Canada
}
\author{Sarah Schon}
\affiliation{%
Arthur B. McDonald Canadian Astroparticle Physics Research Institute, Kingston ON K7L 3N6, Canada
}

\author{Aaron C. Vincent}
\email{aaron.vincent@queensu.ca}
\affiliation{
Department of Physics, Engineering Physics and Astronomy,\\
Queen’s University, Kingston ON K7L 3N6, Canada
}
\affiliation{%
Arthur B. McDonald Canadian Astroparticle Physics Research Institute, Kingston ON K7L 3N6, Canada
}

\begin{abstract}
 The recent discoveries of high-redshift quasars at $z\sim6-10.1$ present a challenge to conventional supermassive black hole (SMBH) formation scenarios: their central SMBHs are too large to have grown from early stellar remnants, even under efficient super-Eddington accretion. An alternative approach is the direct collapse of primordial, dust- and metal-free gas clouds into black hole seeds as early as $z\sim20$. To prevent fragmentation from highly efficient molecular hydrogen cooling and allow such a collapse, a strong flux of UV photons is required to photodissociate molecular hydrogen, suppressing star formation. We investigate whether the decay or annihilation of dark matter with masses below $\sim200$ eV can produce sufficient UV radiation, both from the cosmological background and within early dark matter halos. Using thermal simulations of a one-zone primordial gas cloud model with the Grackle chemistry and cooling library, we map the resulting gas behavior across a range of dark matter lifetimes, cross sections, and masses. From this, we determine regions of parameter space where direct collapse into an SMBH is possible and the corresponding maximum halo mass required.     

\end{abstract}

\maketitle

\section{Introduction}
 Over the past decades, cosmological studies have established a broad theoretical framework for the evolution of the early Universe~\cite{Bertschinger:1998tv, primack1997dark,Gnedin:2000uj,loeb2013first,Dietrich:2001fc}. After recombination, the cosmic medium was dominated by neutral atoms, primarily hydrogen and helium. During this period, the Universe became transparent to visible light, marking the onset of the so-called dark ages~\cite{Miralda-Escude:2003lmt,Jester:2009dw,Parijskij:1998ycu}. With the formation of the first stars, UV radiation reionized the intergalactic medium (IGM) and illuminated the Universe~\cite{Rees:1999kna,Rees:1996ij}. Although multiple critical processes, including the formation of the first stars, occurred during this era, direct observational signatures have long been absent due to technological limitations. New astronomical facilities such as the James Webb Space Telescope (JWST) have opened a new portal for probing the Universe at redshifts up to $z\sim14$~\cite{Robertson:2022gdk,bradley2023high,adams2023discovery,wu2025jades}. Moreover, future 21-cm experiments are expected to extend observational constraints to even higher redshifts, potentially reaching $z\sim30$~\cite{Ciardi:2015oia,HERA:2025ajm,Garsden:2021kdo,Voytek:2013nua}. These developments promise to significantly transform our understanding of cosmic dawn. With the rapid growth of high-redshift observations, a fundamental question has become increasingly critical: what led to the rapid formation of massive astronomical objects, including an X-ray luminous AGN at $z\sim10.1$~\cite{Goulding:2023gqa}, early galaxies~\cite{Menci:2022wia}, and so-called little red dots~\cite{Matthee:2023utn}? Many of these systems host supermassive black holes at their centers, implying that such black holes must have formed at even earlier times. This is in tension with the conventional picture in which SMBHs grow from the remnants of the first stars: sustained Eddington-limited accretion onto such light seeds struggles to explain the early appearance of these SMBHs~\cite{Haiman:2004ve}.

An alternative scenario is the direct collapse black hole (DCBH). In this model, an overdense region arising from a density fluctuation in the early universe becomes a primordial, dust- and metal-free gas cloud as the universe expands. Later, the gas cloud collapses into a black hole seed as early as $z\sim20$~\cite{Tanaka:2008bv}. Normally, such gas clouds would fragment due to efficient cooling by molecular hydrogen, forming stars instead of a single SMBH seed. An efficient mechanism to destroy a significant fraction of H$_2$ is therefore required for monolithic collapse. This can arise from a strong flux of UV photons that either photodissociate molecular hydrogen or photodetach the electron from intermediate H$^-$ ions, in turn suppressing star formation.

In the standard DCBH scenario, the external photon flux would be from a nearby star-forming galaxy whose UV photons must travel to the pristine cloud, where they may suppress the $\rm H_2$ abundance~\cite{Omukai:2000ic, Johnson:2006gd, Dijkstra:2008jk, Shang:2009ij, Agarwal:2012qm, Safranek-Shrader:2012zig, Regan:2014maa, luo2020direct}. Other works have considered sub-regions of the host halo itself, whose star formation may produce the necessary UV radiation to suppress $\rm H_2$ formation elsewhere in the same halo, allowing for more localized direct collpases~\cite{chiaki2023direct, dunn2018sowing}. Aside from the required geometric coincidences, these scenarios imply the first SMBH seed formation would be later than the first star formation during cosmic dawn. Other direct collapse mechanisms that have been proposed have included primordial black holes (PBHs)~\cite{zhang2025novel, pandey2018role, Lu:2023xoi}, turbulence driven direct collapse~\cite{latif2022turbulent}, self-interacting dark matter~\cite{pollack2015supermassive, choquette2019early, grant2025early}, gas streaming velocities~\cite{Tanaka:2013boa, Hirano:2017wbu}, and cosmic strings~\cite{Bramberger:2015kua, Cyr:2022urs}. 

Here, we delve into the possible impact of particle dark matter (DM) on UV injection, direct collapse, and star formation. Many DM models predict DM decay or annihilation to Standard Model (SM) particles, plausibly injecting photons that alter the thermal and ionization history of primordial gas clouds, thereby influencing their evolutionary outcomes.

Several past studies have investigated such effects in the context of early star and SMBH formation. Ripamonti et al.~\cite{Ripamonti:2006gr} studied the effects of sterile neutrinos and light dark matter (LDM) on structure formation, where sterile neutrinos decay into active neutrinos and photons, directly impacting the IGM, while LDM decays and annihilates into electron-positron pairs. Their one-dimensional hydrodynamic simulations demonstrated an increase in gas temperature and an enhancement in coolants ($\rm H_2$ and HD), affecting the thermal and ionization history of the IGM via interactions including photoionization of $\rm H$ and $\rm He$ species and photodissociation of $\rm H_2$ and $\rm HD$. Furthermore, they found that for all DM models they considered, the critical halo mass for the collapse was often higher than in the unperturbed case, indicating that DM decay and annihilation tend to delay structure formation. Friedlander et al.~\cite{Friedlander:2022ovf} investigated the effects of sub-keV DM on SMBH formation, where they assumed that a gas cloud at the DM halo experiences a homogeneous UV background from DM decay/annihilation. In the analysis of the critical curve in the photodetachment ($k_{\rm H^-}$)--photodissociation ($k_{\rm H_2}$) plane, which characterizes the suppression of $\rm H_2$, they found that UV photons produced by sub-keV dark matter within the gas cloud’s halo can suppress molecular hydrogen formation, making SMBH formation via direct collapse more likely, particularly in a large halo. This $k_{\rm H^-}- k_{\rm H_2}$ critical curve method was also used by Jiao et al.~\cite{Jiao:2025kpn}, who showed that the parametric resonant excitation of photons in an ultralight DM halo may produce a sufficient Lyman-Werner (LW) photon flux to prevent the fragmentation of the gas cloud and thereby contribute to SMBH formation. Qin et al.~\cite{Qin:2023kkk} studied the influence of exotic energy injection into the IGM on star formation at early epochs, finding that such injection could either accelerate or delay star formation, depending on the DM properties. Using cooling efficiency as a criterion for star formation, they determined that energy injection from decaying or annihilating DM can either increase or decrease the critical mass or virial temperature threshold required for galaxy formation, depending on the redshift at which halos virialize. Simultaneously, their work extended the mass range of dark matter up to TeV scales by modeling energy cascades injected in homogeneous IGM using methods initially developed for DM searches with the CMB  \cite{Liu:2023fgu,Liu:2023nct}. Recently, Lu et al.~\cite{Lu:2024zwa} studied the suppression of molecular hydrogen by LW photons generated from relic particle decay. Their model included the effects of adiabatic contraction of the dark matter halo and the in-situ production of LW photons within the cloud, which reduces the cloud’s shielding effect. They concluded that considering these effects, dark matter or a short-lived particle $X$ decay could effectively halt $\rm H_2$ production, facilitating SMBH formation via direct collapse. Finally, Lu et al.~\cite{Lu:2023xoi} discussed the possibility of clustered exploding PBHs heating up the baryonic gas cloud, suppressing $\rm H_2$ formation and thus production of SMBH seeds, even for a PBH population that is a subdominant dark matter component.

While each of these works demonstrates that dark matter decay and/or annihilation can impact the evolution of early gas clouds, several gaps remain. For instance, Ripamonti et al.~\cite{Ripamonti:2006gr} did not account for the changes in photodissociation and photodetachment rates induced by energy injection from dark matter, thus rendering the $\rm H_2$ suppression mechanism incomplete. Similarly, the lack of detailed chemical evolution in Friedlander et al.'s~\cite{Friedlander:2022ovf} study limits the understanding of the halo’s thermal evolution and the subsequent formation of SMBHs. Here, we thus seek to fill these gaps by simultaneously modeling photon injection, thermal and chemical evolution, and cloud collapse, to determine in a more robust way how exotic energy injection can affect star and SMBH formation. 

To do so, we construct a toy one-zone model of gas clouds located at the centers of DM halos. We include the photon flux from decaying/annihilating DM in the halo and the background, as well as Cosmic Microwave Background (CMB) photons, as external energy injection into the gas cloud. We then track the thermal and ionization evolution of the gas cloud, including multiple chemical reactions to build a complete chemical network, to study the cloud's evolution from high redshift. 

For this work, we will only consider the influence of DM with masses below $200\ \rm eV$, which produces photons at the low energies (in particular the LW band) needed to suppress the fraction of $\rm H_2$ in the gas cloud. Dark matter models in this mass range are somewhat baroque. While viable models can be built as e.g., a subcomponent of the total DM relic density, they are meant to stand in for more realistic WIMP-scale models in the GeV range and above. We do not tackle these here, since the energy cascade from GeV to eV scale annihilation or decay products is a non-trivial process that has not been fully worked out in such high-redshift clouds. This will be the subject of future work.

This article is structured as follows. In Section~\ref{modeling}, we discuss the modeling of the DM halo and the gas cloud, as well as the photon flux from DM decay/ annihilation in the DM halo and cosmic background; then we outline the relevant gas chemistry and critical conditions for the star/ SMBH formation. In Section~\ref{results} we show the effects of exotic UV injection on a single gas cloud, and in Section \ref{sec:BHMF} we map the resulting seeds onto a black hole mass function and compare with observations. We conclude in Section~\ref{conclusion}.

\section{Modeling}\label{modeling}
\begin{figure}[!ht]
\centering
\includegraphics[width=\columnwidth, pagebox=trimbox]{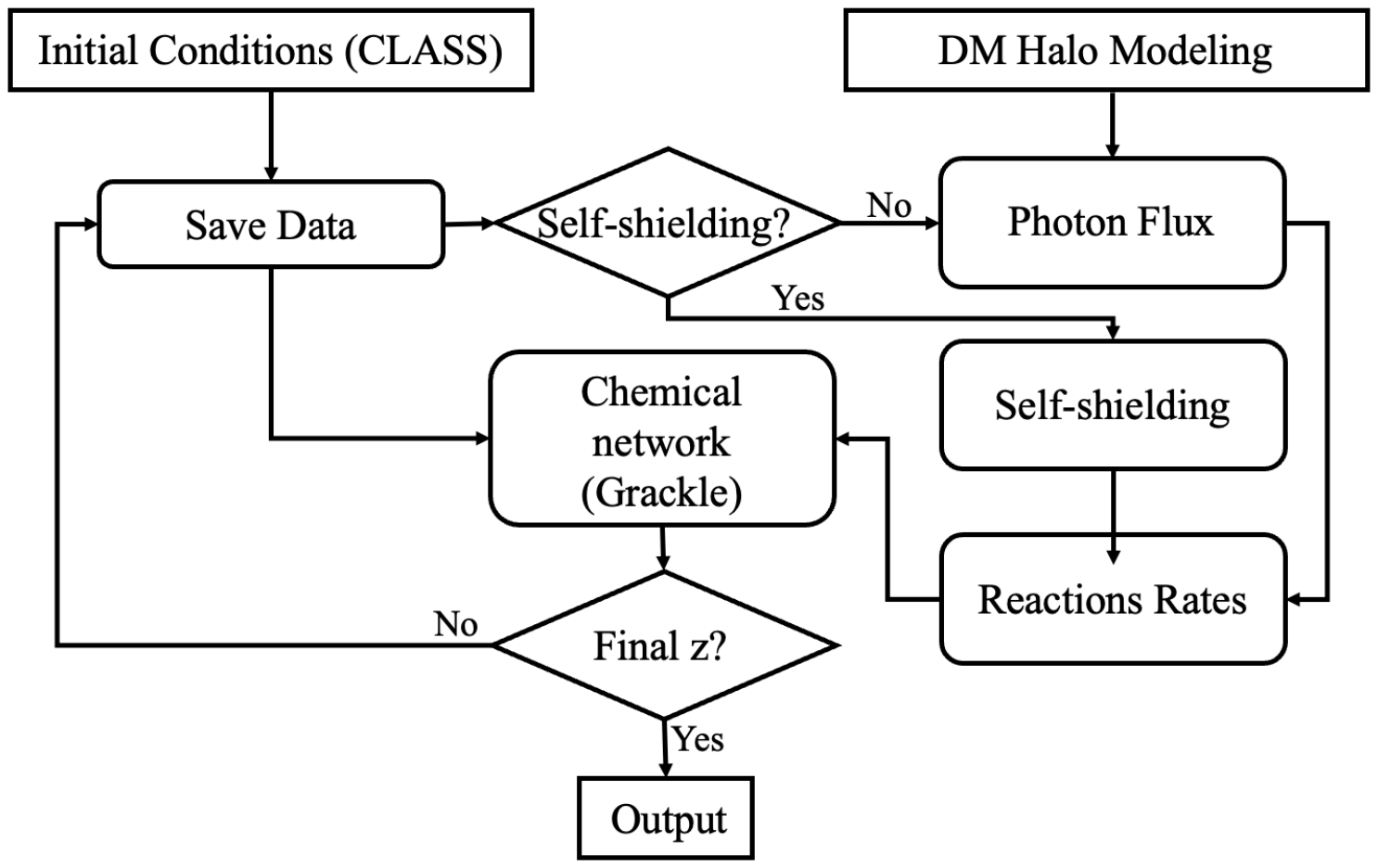}
\caption{\label{fig:flowchart}Flowchart illustrating our method to predict the outcome of gas cloud collapse as affected by UV radiation from dark matter. The corresponding code is available at \url{https://github.com/hanwuqu/OZDM.git}}.

\end{figure}

Our goal is to determine how external photon injection from DM decay and annihilation affects the thermal and chemical evolution and final outcomes of primordial gas clouds. We first construct a one-zone gas cloud model with a mass in the range $10^6-10^7\ \sun$, for which $T_{\rm vir}\sim 10^3\ \rm K$, so that $\rm H_2$ is dominant coolant. We then model the host DM halo using a hybrid method to account for the transition from overdensity to halo, which we will detail in Section~\ref{DMHalo}. The total photon flux from DM decay/annihilation is subsequently computed. This photon flux, together with the gas cloud profile, is used as input to the thermal and chemical evolution code \texttt{Grackle}~\cite{Smith:2016hsc} to obtain the thermal and ionization history of the gas cloud without photon self-shielding. Additional calculations involving the fractions of different H and He species in the gas cloud are then required to obtain the photon self-shielding fraction, which is subsequently used to update the thermal and ionization evolution. This overall pipeline is illustrated in Fig.~\ref{fig:flowchart}.

\subsection{Dark Matter Halo Modeling}\label{DMHalo}
Photon injection from dark matter annihilation or decay will come from two distinct sources: 1) the dark matter halo surrounding the gas cloud itself, and 2) the homogeneous background contribution, which we will return to in \ref{sec:fluxes}. Since the dark matter-induced UV flux from a given point in the halo scales as density for decay and density squared for annihilation, the choice of dark matter density profile can significantly affect the predicted flux. Because we are evolving from near recombination to the epoch of reionization, we require a smooth parametrization for the DM halo throughout its entire history. Well-resolved halos in DM/hydro simulations are only available for redshifts well past those of interest here. We therefore adopt a hybrid method that uses a standard Navarro-Frenk-White (NFW) halo shape~\cite{navarro1997universal}, with an empirical parametrization for the concentration parameter redshift evolution, and a simplified shell-crossing model to account for the physical halo size. 

The standard NFW profile can be written
\begin{equation}
    \frac{\rho(r)}{\rho_{\rm crit}}=\frac{\delta_c}{(r/r_s)(1+r/r_s)^2},
\end{equation}
 where $r_s$ is the scale radius and  $\delta_c$ is a characteristic dimensionless density, 
\begin{equation}
    \delta_c=\frac{200}{3}\frac{c^3}{\left[\ln\left(1+c\right)-c/\left(1+c\right)\right]},
\end{equation}
and $\rho_{\rm crit}$ is the critical density of the Universe. The concentration parameter $c$ is defined as
\begin{equation}
    c\equiv\frac{r_{200}}{r_s}, 
\end{equation}
where $r_{200}$ is the radius of a sphere that encloses an average density 200 times  $\rho_{\rm crit}$. $c$  depends on the mass of the halo as well as the redshift, and can be seen as governing the evolution of halos as they form and contract. 

While high-resolution results of large-scale simulations exist at redshifts of a few, the evolution of $c(M,z)$ at higher redshifts remains unsettled. Early studies found that halo concentrations decrease with redshift and exhibit considerable scatter at fixed mass~\cite{jing2000density,bullock2001profiles}. Zhao et al.~\cite{Zhao:2002rm} incorporated the mass accretion history (MAH) of halos and showed that halo growth generally consists of an early fast-accretion phase and a late slow-accretion phase, separated at $c\sim4$. This transition has been widely adopted in subsequent halo evolution studies~\cite{zhao2003mass,Li:2005jm,Hoffman:2007tu}. We therefore adopt the following phenomenological parametrization, aimed at smoothly interpolating between a weakly concentrated profile and a more evolved $c = 4$ halo: 
\begin{equation}
    c=\left\{
\begin{aligned}
& 1+3e^{-\left(z-z_{\rm vir, halo}\right)}\quad &z\ge z_{\rm vir, halo} \\
& 4\quad &z_{\rm final}\le z\le z_{\rm vir, halo} \\
\end{aligned}
\right.,
\end{equation}
where $z_{\rm vir, halo}$ is the virial redshift of the DM halo, $z_{\rm final}$ is the end of simulation, which we set to $z_{\rm final}=10$ to coincide with the reionization epoch of the Universe is $z\sim6-15$~\cite{Zaroubi:2012in, Wise:2019qtq}.

Finally, the evolution of the halo radius $r_{200}$ is required to parametrize the physical halo size. To model this, we employ a shell-crossing model that smoothly tracks the initial linear evolution through its gravitational collapse
~\cite{baumann2022cosmology}. Ref.~\cite{Lu:2023xoi} provides a simplified model of halo evolution based on this method. We track the outermost shell $r = r_{200}$:
\begin{equation}
\frac{d^2r}{dt^2}=-\frac{GM_{halo}}{r^2},
\end{equation}
where $\rm M_{halo}$ is the mass of DM halo and $G$ is the gravitational constant. At very high redshift (e.g., $z\sim 1100$), before the overdensity becomes nonlinear, the mass shell is almost moving with the Hubble flow, but with a correction, such that the initial velocity of the mass shell is 
\begin{equation}
    v_{phy}=Hr_{phy}+v=Hr_{phy}\left(1-\frac{\delta}{3}\right),
\end{equation}
where $v_{\rm phy}$ and $r_{\rm phy}$ are respectively the physical velocity and radius of the mass shell and $H$ is the Hubble parameter and $\delta\equiv\frac{\rho_{\rm halo}}{\rho_{\rm crit}}-1$ is the overdensity. With this definition, the initial radius $r_{\rm ini}$ can be obtained from 
\begin{equation}
    M_{\rm halo}=\frac{4}{3}\pi r_{\rm ini}^3\delta\rho_{\rm crit}.
\end{equation}
Therefore, the only free parameter in this model is the overdensity $\delta$. The CMB tells us that the RMS fluctuations are of the order $\delta\sim10^{-5}$. However, as we are interested in the earliest halos to virialize, we take $\delta=0.02$ to ensure the halo has virialized by redshift $\sim 15$. This relatively high $\delta$ implies that DCBH and early star formation at high redshift are rare.

\subsection{Gas Cloud Modeling}
To model the evolution of the gas cloud, we adopt a one-zone model, assuming a uniform gas cloud to describe its density evolution. Ref.~\cite{Tegmark:1996yt} provides a fitting function for the density evolution of a one-zone gas cloud,
\begin{equation}
\rho_{\rm gas\ cloud}(z)=\rho_{crit}\left(1+z\right)^3\exp\left(\frac{1.9A}{1-0.75A^2}\right),
\end{equation}
where
\begin{equation}
A=\frac{1+z_{\rm vir}}{1+z},
\end{equation}
and $z_{\rm vir}$ is the virial redshift of the gas cloud. This function was used to numerically integrate a system of chemical equations and determine the minimum mass of a virialized gas cloud required to cool within a Hubble time, yielding results comparable to those from computationally expensive 3D simulations. At the early stage, the gas cloud is diluted by the expansion of the Universe. Then the gas cloud becomes concentrated and collapses when the redshift approaches $z_{\rm vir}$, where the density of the gas cloud reaches the virial density
\begin{equation}
\rho_{\rm vir}=18\pi^2\rho_{\rm crit}\left(1+z\right)^3.
\end{equation}
After virialization, we assume that the density of the gas cloud remains invariant until the end of the simulation. 

With the evolution of the gas cloud, chemical reactions and adiabatic work contribute to the alteration of its internal energy. Therefore, the temperature drops before the gas cloud reaches virialization, as adiabatic expansion dominates. The temperature then rises when the gas cloud approaches virialization. However, in a realistic three-dimensional gas cloud, the temperature increases dramatically due to shock heating. This effect cannot be obtained from a one-zone model because of the absence of the spatial density distribution of the gas cloud. To account for shock heating, we therefore raise the temperature of the gas cloud at $z_{\rm vir}$ to the virial temperature $\rm T_vir$. Ref.~\cite{Barkana:2000fd} provides a fitting function for the virial temperature 
\begin{equation}
\label{eqs:Tvir}
    T_{\rm vir}\approx244~\rm K\left(\frac{\Omega_m h^2}{0.14}\right)^{1/3}\left(\frac{M_{cloud}}{10^4 M_\odot}\right)^{2/3}\left(\frac{1+z_{vir}}{100}\right),
\end{equation}
where $\Omega_m$ is the matter density parameter, $h$ is defined from $H_0=100h\rm km\cdot s^{-1}\cdot Mpc^{-1}$ where $H_0=67.36\ \rm km\cdot s^{-1}\cdot Mpc^{-1}$ is the Hubble parameter today. Although its density remains invariant after gas cloud virialization, thermal evolution continues because the chemical components absorb/release heat and further change the temperature of the gas cloud. We track the thermal history of the gas cloud after virialization until the beginning of reionization at $z<10$, then categorize the gas cloud evolution outcomes according to the criteria of different structure formation based on the cooling behavior of the gas cloud. These will be described in Section~\ref{criteria}.

\subsection{Photon Flux from DM Decay/ Annihilation} \label{sec:fluxes}
The thermal evolution of the gas cloud after virialization largely determines its fate, and it is set by the coolants and cooling channels available in the gas. In the clouds we consider, the virial temperature is below $\sim10^4$~K, where atomic hydrogen cooling is inefficient, so molecular hydrogen is the dominant coolant. The $\rm H_2$ abundance is in turn sensitive to external radiation: photons can destroy $\rm H_2$ directly through photodissociation, or suppress its formation by photodetaching $\rm H^-$, the key intermediate in the main formation channel. Because DM makes up most of the matter in the Universe, its decay or annihilation products are a natural source of such photons. In this work, we therefore treat photons from DM decay or annihilation, both The flux of external photon injection is
\begin{equation}
\label{flux}
J\left(\vec{r}, E\right) = \frac{E}{4\pi}\int dV'\frac{dn_\gamma}{dEdt}\left(\vec{r}, E\right)\frac{1}{\left(\vec{r'}-\vec{r}\right)^2},
\end{equation}
where $dn_\gamma/dEdt$ is the differential injection rate of photons from dark matter decay or annihilation per unit volume. Since the gas cloud is located at the center of the halo, and the photon flux from dark matter decay or annihilation is evaluated at this center, the gas cloud also serves as the origin of the integration. Therefore, Eq.~(\ref{flux}) can be simplified to
\begin{equation}
\label{flux_gen}
    J\left(E\right)=E\int dr\frac{dn_\gamma}{dEdt}.
\end{equation}  
This flux must be redshifted from the time of injection to the time of deposition $z_{\rm obs}$ in the gas cloud, namely, 
\begin{equation}
    \frac{dn_\gamma}{dEdt}=\frac{\left(1+z_{\rm obs}\right)^3}{\left(1+z\right)^3}\frac{dn_\gamma'}{dE'dt'},
\end{equation}
where primed quantities refer to values at the emission source at redshift $z$, while unprimed symbols denote quantities evaluated at the gas cloud at redshift $z_{\rm obs}$. In terms of redshift, this becomes
\begin{equation}
\label{flux_bg}
    J_{\rm bg}(E)=\frac{\left(1+z_{obs}\right)^3}{4\pi}\int_{z_{obs}}^\infty\frac{dz}{H(z)\left(1+z\right)^3}E\frac{dn_\gamma'}{dE'dt'}.
\end{equation}
With the geometry and time dependence of energy deposition accounted for, we now turn to the photon spectra at injection.
\subsubsection{Dark Matter Decay}
\begin{figure}[!ht]
\centering
\includegraphics[width=\columnwidth, pagebox=trimbox]{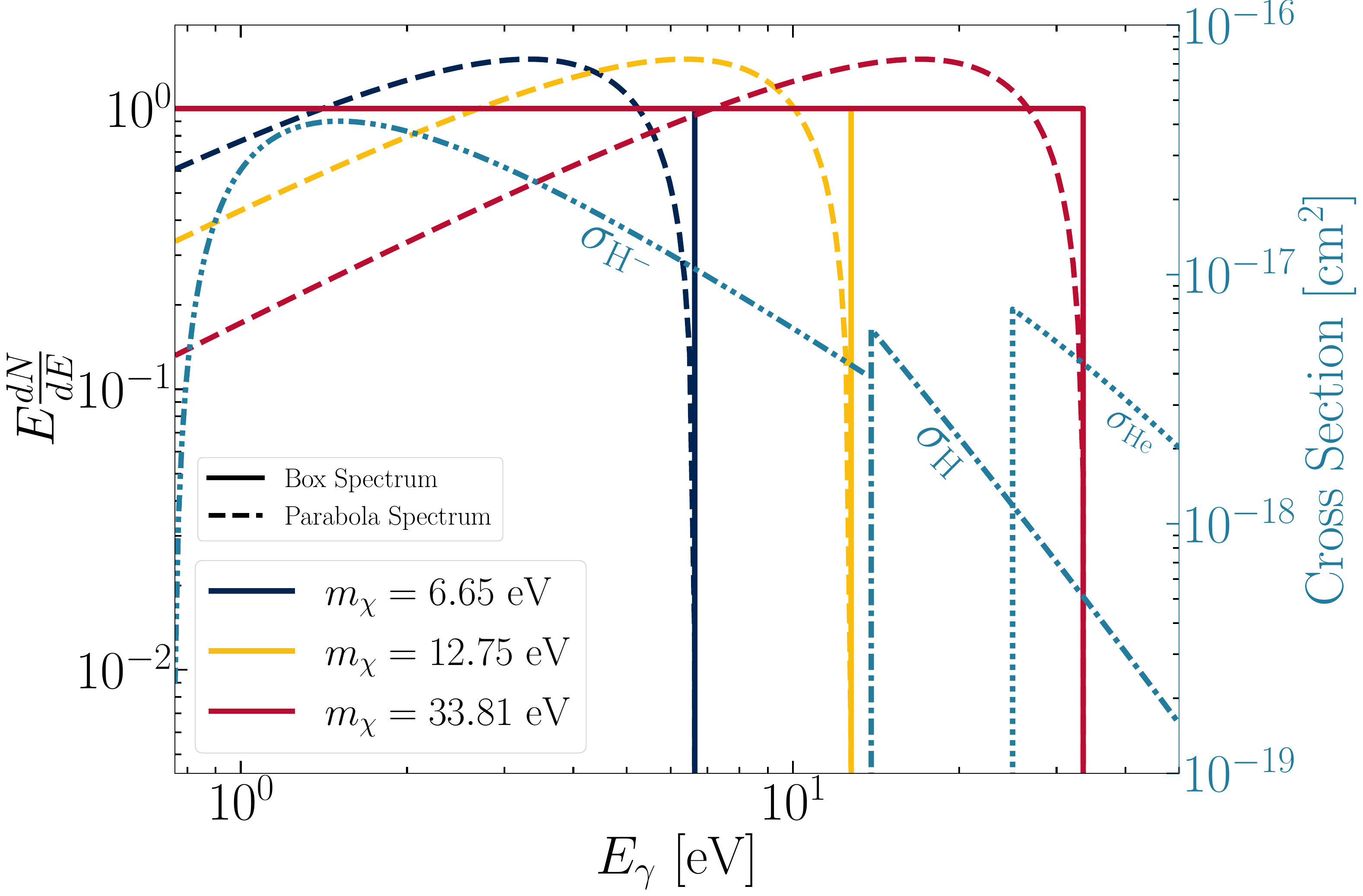}
\caption{\label{fig:spectra}Box (solid) and parabola (dashed) spectra of photon flux from DM with $m_\chi=6.65,\ 12.75,\ 33.81\ \rm eV$. Cerulean curves correspond to photodetachment ($\sigma_\mathrm{H^-}$, dash-dot-dotted), photoionization of $\rm H$ ($\sigma_\mathrm{H}$ dash-dotted) and $\rm He$ ionization ($\sigma_\mathrm{He}$ dotted) cross sections.} 

\end{figure}

In order to avoid depleting the DM abundance, the lifetime $\tau$ must be much larger than the age of the Universe $t_0 \sim 10^{17}$s, or constitute only a small fraction $f_{dm}$ of the total CDM component. We therefore write the in-situ photon injection rate from decaying DM as
\begin{equation}
    \frac{dn_\gamma}{dEdt}=\frac{f_\gamma f_{dm}\rho_{dm}(r)}{\tau m_{\chi}}\frac{dN}{dE}(E),
\end{equation}
 where $m_{\chi}$ and $\tau$ are the mass and lifetime of the dark matter, respectively. $f_\gamma$ is the fraction of energy released as photons in each decay, and $\rho_{dm}$ is the density of the dark matter halo. The spectral distribution of injected photons, $dN/dE$, depends on the final-state phase space resulting from dark matter decay. Although a two-body final-state decay is intuitively expected, both photodissociation and photodetachment reactions require photons in the LW range; therefore, a continuous spectrum is desired. We consider two  representative spectra: a box-shaped photon spectrum and an $N$-body final state with $N>2$. The box-shaped photon spectrum is
\begin{equation}
    \left.E\frac{dN}{dE}\right|_{\rm box}=\Theta\left(1-\frac{E}{m_{\chi}}\right),
\end{equation}
where $\Theta\left(x\right)$ is the Heaviside step function. As described in Ref.~\cite{Ibarra:2012dw}, this can be realized through decay to a pair of mediators, such as light scalar particles, which further decay into photons. The photon spectrum from an $N$-body final state ($N>2$) is continuous, as required by those chemical reactions. Nevertheless, due to the dependence on the exact number of final-state particles, the N-body final-state spectrum can be approximated by a parabola-shaped curve~\cite{Friedlander:2022ovf}.
\begin{equation}
    \left.E\frac{dN}{dE}\right|_{\rm parabola}=6\left[\frac{E}{m_{\chi}}-\left(\frac{E}{m_{\chi}}\right)^2\right]\Theta\left(1-\frac{E}{m_{\chi}}\right).
\end{equation}
The shapes of both spectra for photons from $m_\chi=6.65,\ 12.75$ and 33.81 eV DM are shown in Fig.~\ref{fig:spectra} as solid and dashed lines. Comparing with the box-shaped spectrum, which is almost flat in the whole energy range, the parabola-shaped spectrum contains a bump at photon energy $E_\gamma$ near the energy threshold.
The photon flux produced from DM decay in the DM halo is
\begin{equation}
    J_{\rm decay, halo}(E)=\frac{f_\gamma f_{\rm dm}}{\tau m_\chi}E\frac{dN}{dE}\int dr'\rho_{\rm halo}(r'),
\end{equation}
where $\rho_{\rm halo}$ is the NFW profile density obtained from the hybrid method described in Sec.~\ref{DMHalo}. Adopting the box and parabola photon spectra into the Eq.~\ref {flux_bg}, the photon flux from the background dark matter decay is
\begin{equation}
\begin{aligned}
    J_{\rm decay,\ bg}(E, z_{obs})=&\frac{(1+z_{obs})^3}{4\pi}\frac{f_\gamma f_{\rm dm}\Omega_{\rm dm}\rho_{crit}}{\tau m_{\chi}}\times\\
    &\int_{z_{obs}}^{z_{max}}\frac{dz}{H(z)},
\end{aligned}
\end{equation}
for the box spectrum, where $\Omega_{dm}$ is the density parameter of dark matter, $\rho_{crit}$ is the critical density of the universe at present, $z_{obs}$ is the redshift at which the gas cloud is observed and $z_{max}=m_{\chi}/(1+z_{obs})E-1$ is the maximal redshift deduced from the Heaviside step function. And 
\begin{equation}
\begin{aligned}
    &J_{decay,\ bg}(E, z_{obs})=\frac{3}{2\pi}\frac{f_\gamma f_{dm}\Omega_{dm}\rho_{crit}}{\tau m_{\chi}}\times\\
    &\int_{z_{obs}}^{z_{max}}dz\left[\frac{E}{m_{\chi}}\frac{(1+z)(1+z_{obs})^2}{H(z)}\right.\\ &\left.-\frac{E^2}{m_{\chi}^2}\frac{(1+z)^2(1+z_{obs})}{H(z)}\right]
\end{aligned}
\end{equation}
is the photon flux with the parabola spectrum for the background DM decay.
\subsubsection{Annihilation}
For s-wave annihilating dark matter, the photon injection rate is
\begin{equation}
    \frac{dn_\gamma}{dEdt}\left(E, r\right)=\frac{f_\gamma f_{dm}^2\rho^2_{dm}\left(r\right)\langle \sigma v\rangle}{m^2_{\chi}2^p}\frac{dN}{dE}\left(E\right),
\end{equation}
 where $\langle\sigma v\rangle$ is the thermally averaged annihilation cross section of the dark matter, $p = 1$ for self-conjugate dark matter (e.g., a Majorana fermion) and $p = 2$ for dark matter which consists of a particle and antiparticle (e.g., Dirac). We will present results using the parabola spectrum defined above. The photon flux from dark matter annihilation in the halo is
 \begin{equation}
 \begin{aligned} 
     J_{ann, halo}(E, z)=&\frac{6f_\gamma f_{dm}^2\langle\sigma v\rangle}{2^p m_{\chi}^2}\int dr'\rho_{\rm halo}^2(r')\\
     &\left[\frac{E}{m_{\chi}}-\left(\frac{E}{m_{\chi}}\right)^2\right]\Theta\left(1-\frac{E}{m_{\chi}}\right).
\end{aligned}
\end{equation}
Similarly, the photon flux from dark matter annihilation from the background is 
\begin{equation}
\begin{aligned}
     &J_{ann, bg}=\frac{3}{2\pi}\frac{f_\gamma f_{dm}^2\Omega_{dm}^2\rho_{cirt}^2\langle\sigma v\rangle}{2^p m_{\chi}^2}
     \\
     &\int_{z_{obs}}^{z_{max}}dz\left[\frac{(1+z_{obs})^2E}{m_{\chi}}\frac{(1+z)^4}{H(z)}-\frac{(1+z_{obs})E^2}{m_{\chi}^2}\frac{(1+z)^5}{H(z)}\right].
\end{aligned}
\end{equation}

\subsection{Evolution of Thermal and Ionization History of the Gas Cloud}

We use the \texttt{Grackle}~\cite{Smith:2016hsc} package to track the full thermal and chemical evolution simulation, accounting for reactions between different ionization states of H and He, electrons and photons. As a chemistry and cooling library, \texttt{Grackle} provides the treatment of non-equilibrium chemistry and cooling of primordial gas by solving a series of equations to update the chemical network and evolving the Lagrangian energy equation to take radiative cooling and heating processes into account within a specific time range during the evolution of the gas cloud density.\\
We obtain the initial conditions for the simulation at $z\sim1000$ from \texttt{CLASS} (Cosmic Linear Anisotropy Solving System)~\cite{Lesgourgues:2011rg} including the temperature and the fractions of $\rm H^+$ and $\rm He^+$. The abundance of $\rm He$ is set to $\rm Y_{He}=0.24$ and the fractions of $\rm H^-$ and $\rm He^{++}$ are set to a small value close to zero.
With these initial values, we use \texttt{Grackle} to evolve the gas cloud with density change $\Delta \rho$ over the redshift interval $\Delta z$, then collect outputs such as the temperature and number densities of species as the input for the next redshift step.

In \texttt{Grackle}, the primordial gas contains nine species ($\rm H$: $\rm H$, $\rm H^+$, $\rm H^-$, $\rm H_2$, $\rm H_2^+$; $\rm He$: $\rm He$, $\rm He^+$, $\rm He^{++}$, and $e^-$) involved in 31 reactions. Although most of them are pre-built in the package, eight reactions need to be specified manually due to their dependence on the external photon flux injection. The reaction rate in the optically thin region can be estimated from
\begin{equation}
\label{rate}
    k_i=\int d\nu\frac{4\pi J_\nu}{h\nu}\sigma_i(\nu),
\end{equation}
where $\nu$ is the frequency of the photon, $J_\nu$ is the flux of injected photons, and $\sigma_i$ is the cross section of the reaction, which we provide in Appendix~\ref{chemical reactions}. In the mass range of gas clouds considered in this study, the virial temperature is below $\sim10^4 \rm K$~\cite{barkana2001beginning} and the dominant coolant that may affect the evolutionary outcome is molecular hydrogen. Since a moderate cooling rate is required to prevent fragmentation of the gas cloud and allow SMBH seed formation, we focus on reactions that directly destroy or suppress the formation of molecular hydrogen. The dominant $\rm H_2$ formation mechanism at number densities below $\sim 10^{3}\ \rm cm^{-3}$ is a two-step process:
\begin{equation}
\begin{aligned}
    \rm H +e^-\rightarrow H^-+\gamma,\\
    \rm H+H^-\rightarrow H_2+e^-. \label{eq:H2formation}
\end{aligned}    
\end{equation}
Other $\rm H_2$ formation channels are via three-body processes: $\rm H+H+H\rightarrow H_2 +H$ and $\rm H+H+H_2\rightarrow H_2+H_2$, which are much rarer. Hence, there are two $\rm H_2$ suppression mechanisms, the direct destroy of $\rm H_2$ by \textit{photodissociation}:
\begin{equation}
    \rm H_2+\gamma\rightarrow H+H,
\end{equation}
and the indirect suppression of $\rm H_2$ formation by reducing the reactant, which is known as \textit{photodetachment}:
\begin{equation}
    \rm H^-+\gamma\rightarrow H+e^{-}.
\end{equation}
Photodissociation requires photons in the LW energy range of $11.2-13.6\ \rm eV$, while photodetachment adapts to a broader energy range of $0.76\ \rm eV\le E_\gamma\le 13.6\ \rm eV$ photons. Thus, the suppression of $\rm H_2$ in the gas cloud not only depends on the intensity but also on the shape of the spectrum of UV flux. The reaction rate of photodetachment can be obtained from the convolution of the photodetachment cross section with the flux of photons as Eq.~\ref{rate}, where the cross section can be estimated as~\cite{shapiro1987hydrogen}
\begin{equation}
    \sigma_{\rm H^-}\approx4.31\times 10^{-18}\frac{\tilde{E}^3}{\left(0.0555+\tilde{E}^2\right)^3}\ \rm cm^2,
\end{equation}
where
\begin{equation}
    \tilde{E}=\sqrt{\frac{E_\gamma-0.754\ \rm eV }{13.6\ \rm eV}}.
\end{equation}
Unlike photodetachment and photoionization reactions, the cross section of photodissociation is highly discontinuous because of the multiple excited states of molecular hydrogen, thus an approximation formula is used to estimate the photodissociation rate~\cite{wolcott2017beyond},
\begin{equation}
    k_{\rm H_2}\approx1.39\times10^{-12}\mathrm{s}^{-1}\left(\frac{J_{\rm LW}}{10^{-21}\ \rm erg \ s^{-1}\ Hz^{-1}\ cm^{-2}\ sr^{-1}}\right),
\end{equation}
where $J_{\rm LW}$ is the flux of LW photons. All of the reactions that depend on external photon injection have a linear dependence on the flux of UV photons, therefore, with the same spectrum, a strong intensity of flux will cause more aggressive suppression of $\rm H_2$ in the gas cloud. \textit{Photoheating} occurs when the energy of ionizing photon exceeds the ionization threshold of atoms, heating up the surrounding material. This also contributes to the thermal evolution of the gas cloud, especially when the temperature of the gas falls below $\sim 10^4\ \rm K$. This rate can be calculated from the photoionization cross section as well as the UV flux~\cite{Haardt:2011xv}
\begin{equation}
    \mathcal{H}_i=\int d\nu\frac{4\pi J_\nu}{h\nu}h\left(\nu-\nu_i\right)\sigma_i(\nu),
\end{equation}
where $i\in\left\{\rm H, He, He^+\right\}$, $\nu_i$ is the threshold frequency of ionization of corresponding species. It should finally be highlighted that, as a free electron is required in the process \eqref{eq:H2formation}, additional photoionization can also catalyze the \textit{formation} of H$_2$. \\

For all of the reactions above that depend on the UV photon flux, as well as the reactions built into \texttt{Grackle}, the incident radiation has three components: photons from DM decay or annihilation within the halo, the same contribution from the cosmological background, and the CMB, which we model as a blackbody at temperature $T_0(1+z)$ with $T_0 = 2.725$~K. We pass the total photon flux to \texttt{Grackle}, which evolves the chemical network over each redshift step and tracks the species abundances and the temperature of the gas cloud.
\subsection{Self-shielding}
\label{sec:selfshielding}
The gas cloud is exposed to an external photon flux, and the photochemical rates that drive its thermal and chemical evolution depend on how much of that flux actually reaches the gas. As the radiation propagates inward, it is attenuated by absorption in the gas itself. This self-shielding can strongly suppress the effective photo-rates in dense regions, so neglecting it would overestimate the photon injection into the cloud. Since we treat the gas cloud with a one-zone model, the attenuation has to be included in an approximate way.

In principle, self-shielding can be computed from a three-dimensional treatment of the radiative transfer through the cloud; 
In practice, this requires evaluating multidimensional integrals over the photon flux at every step of the simulation, which is prohibitively expensive for a coupled thermal and chemical evolution. The approach also handles $\rm H_2$ poorly. Lyman--Werner photodissociation proceeds through absorption in discrete lines, so the shielding depends on the line structure of the cross section rather than on a smooth continuum opacity. We therefore adopt a simpler strategy: for each relevant reaction we compute the optically thin rate and multiply it by a shielding factor taken from fitting formulae calibrated on detailed radiative transfer calculations. Below we describe this treatment for $\rm H_2$ photodissociation and $\rm HI$ photoionization.

\paragraph{$\rm H_2$ photodissociation.}
Following Ref.~\cite{Wolcott-Green:2011tul}, we write the shielded photodissociation rate as
\begin{equation}
    k_{\rm H_2}\left(N_{\rm H_2}, T\right)=k_{\rm H_2}\left(N_{\rm H_2}=0, T\right)\,f_{\rm sh}\left(N_{\rm H_2}, T\right),
\end{equation}
where $k_{\rm H_2}(N_{\rm H_2}=0, T)$ is the optically thin rate, $N_{\rm H_2}=n_{\rm H_2}L_{\rm Jeans}$ is the column density of $\rm H_2$, $L_{\rm Jeans}$ is the Jeans length and $f_{\rm sh}$ is the self-shielding factor,
\begin{eqnarray}
\label{fsh}
    f_{\rm sh}\left(N_{\rm H_2}, T\right)&=&\frac{0.965}{\left(1+x/b_5\right)^{1.1}}+\\ &&\frac{0.035}{\left(1+x\right)^{0.5}}\exp\left[-8.5\times 10^{-4}\left(1+x\right)^{0.5}\right].\nonumber
\end{eqnarray}
Here $x\equiv N_{\rm H_2}/(5\times10^{14}\ {\rm cm^{-2}})$ and $b_5\equiv b/(10^5\ {\rm cm\,s^{-1}})$, with $b$ the Doppler broadening parameter. Through $b$, the factor depends on the gas temperature as well as on the $\rm H_2$ column density. This is because thermal broadening sets how quickly the individual Lyman--Werner lines saturate.

\paragraph{$\rm HI$ photoionization.}
Once the incident photon energies exceed the $\rm HI$ ionization threshold, the large abundance of neutral hydrogen in the cloud makes photoionization of atomic hydrogen important, and here too, shielding must be taken into account. Self-shielding of $\rm HI$ photoionization has not been studied directly in the environments we consider. However, it has been characterized in cosmological simulations of the intergalactic medium during and after reionization, which give a practical calibration~\cite{Rahmati:2012rg,Chardin:2017juq}. Rahmati et al.~\cite{Rahmati:2012rg} post-processed cosmological hydrodynamical simulations with radiative transfer of the ultraviolet background (UVB) and the recombination radiation. They found that the transition between highly ionized and self-shielded gas is smoother than a sharp density threshold would imply. They gave the following fit for the effective photoionization rate as a function of hydrogen number density:
\begin{equation}
\label{Gamma_ssh}
    \frac{\Gamma_{\rm phot}}{\Gamma_{\rm UVB}}=(1-f)\left[1+\left(\frac{n_{\rm H}}{n_0}\right)^{\beta}\right]^{\alpha_1}+f\left(1+\frac{n_{\rm H}}{n_0}\right)^{\alpha_2},
\end{equation}
where $\Gamma_{\rm UVB}$ is the optically thin rate set by the background, $\Gamma_{\rm phot}$ is the rate at density $n_{\rm H}$, $f$, $\alpha_1$, $\alpha_2$ and $\beta$ are fitting parameters, and $n_0$ is the characteristic density above which self-shielding sets in. The ratio $\Gamma_{\rm phot}/\Gamma_{\rm UVB}$ therefore plays the same role for $\rm HI$ as $f_{\rm sh}$ does for $\rm H_2$ in Eq.~\eqref{fsh}. Chardin et al.~\cite{Chardin:2017juq} confirmed that this functional form also holds at $z\sim6$--$10$ and gave the corresponding fit parameters, which we adopt here.

The characteristic density $n_0$ follows from asking when the cloud becomes optically thick to ionizing photons. For gas in approximate hydrostatic equilibrium, the total hydrogen column density can be estimated as
\begin{equation}
\label{NH}
    N_{\rm H}\sim 1.6\times10^{21}\ {\rm cm^{-2}}\, n_{\rm H}^{1/2}\,T_4^{1/2}\left(\frac{f_g}{0.17}\right)^{1/2},
\end{equation}
with $n_{\rm H}$ in ${\rm cm^{-3}}$, $T_4\equiv T/10^4\ {\rm K}$, and $f_g=\Omega_b/\Omega_m$ the cosmic baryon fraction. Refs.~\cite{Rahmati:2012rg,Chardin:2017juq} relate this to the neutral column under the assumption of highly ionized gas. That assumption does not hold in our cloud, so we instead relate the two directly through the neutral fraction $x_{\rm HI}\equiv n_{\rm HI}/n_{\rm H}$,
\begin{equation}
    N_{\rm HI}=x_{\rm HI}\,N_{\rm H}.
\end{equation}
Self-shielding becomes significant once the optical depth at the ionization edge reaches unity, $\tau_{\rm HI}\simeq1$, which corresponds to
\begin{equation}
\label{NHI}
    N_{\rm HI,\,ssh}\sim4\times 10^{17}\ {\rm cm^{-2}}\left(\frac{\bar{\sigma}_{\rm HI}}{2.49\times10^{-18}\ {\rm cm^2}}\right)^{-1},
\end{equation}
where
\begin{equation}
    \bar{\sigma}_{\rm HI}\equiv\frac{\int^{\nu_{\rm HeII}}_{\nu_{\rm HI}}\left(J_\nu/\nu\right)\sigma_{\rm HI}\,d\nu}{\int^{\nu_{\rm HeII}}_{\nu_{\rm HI}}\left(J_\nu/\nu\right)d\nu}
\end{equation}
is the grey absorption cross section, averaged over the incident spectrum $J_\nu$. Setting $N_{\rm HI}=N_{\rm HI,\,ssh}$ in Eqs.~\eqref{NH}--\eqref{NHI} and solving for $n_{\rm H}$ gives the characteristic self-shielding density
\begin{equation}
\begin{aligned}
    n_0\simeq6.25\times10^{-8}\ {\rm cm^{-3}}\;x_{\rm HI}^{-2}\,&T_4^{-1}\left(\frac{\bar{\sigma}_{\rm HI}}{2.49\times10^{-18}\ {\rm cm^2}}\right)^{-2}\times\\
    &\left(\frac{f_g}{0.17}\right)^{-1}.
\end{aligned}
\end{equation}
Because $x_{\rm HI}$ and $T$ evolve along with the cloud, we update $n_0$, and hence the shielding factor in Eq.~\eqref{Gamma_ssh}, at every step of the simulation.

\subsection{Outcomes}\label{criteria}
The precise criteria for structure formation are not yet fully established, especially for SMBH seed formation. However, in general, star formation requires relatively rapid cooling to trigger fragmentation, whereas SMBH seed formation requires moderate, more uniform cooling. Therefore, we adopt a criterion established in Ref.~\cite{Qin:2023kkk} based on the cooling efficiency. According to this criterion and simulation results, we categorize the possible outcomes of gas cloud evolution as follows:
\begin{description}
    \item[Star Formation] Ref.~\cite{Tegmark:1996yt}  proposed a simple criterion that we will use as a threshold for fragmentation and collapse: we consider a gas cloud to be cooling sufficiently quickly to form stars if
\begin{equation}
T\left(\eta z\right)\le\eta T(z),
\end{equation}
where $\eta$ is chosen to be 0.75. This choice corresponds to a significant temperature drop amount within one Hubble time, allowing collapse and formation of the first stars. We will assume that star formation occurs at the redshift when the above condition becomes an equality.
\item[SMBH Seeds Formation] 
Uniform global cooling of the gas cloud is necessary for the formation of a massive SMBH, which in turn needs to be slow enough to prevent fragmentation. Hence, a gas cloud after evolution may form an SMBH seed if
\begin{equation}
    T\left(\eta z\right)>\eta T(z).
\end{equation}
The gas cloud is further required to satisfy the collapse condition: the mass of gas cloud must exceed the Jeans mass~\cite{jeans1902stability}
\begin{equation}
\label{Jeans}
     M_{\rm gas} \ge M_{J},
\end{equation}
where
\begin{equation}
    \rm M_{J}=\frac{4\pi}{3}\rho_{\rm gas}R^3_{J},
\end{equation}
with $\rm R_J=1/2\lambda_J$ and $\lambda_J$ is the Jeans length
\begin{equation}
     \lambda_J\equiv\frac{c_s}{\sqrt{G\rho}}=\sqrt{\frac{k_BT}{G\mu m_p\rho}}.
\end{equation}
$\mu$ is the mean molecular weight, $m_p$ is the mass of proton, $\rho$ is the density of the cloud and\begin{equation}
    c_s=\sqrt{\frac{k_BT}{\mu m_p}},
\end{equation}
is the speed of sound in the gas cloud.
\item[High Temperature Cloud (HTC) Formation] In our gas cloud model, the density of the gas cloud remains invariant after virialization. This allows the gas cloud to cool without energy changes due to adiabatic contraction/expansion\footnote{In \texttt{Grackle}, this part of the energy is defined as the internal energy of the system; therefore, constant density implies that the internal energy remains invariant}. However, if the fraction of coolant, especially $\rm H_2$, is extremely low, the temperature drop of the gas cloud can be insignificant, and the gas cloud essentially becomes a relatively hot region in the Universe. In this scenario, we require the temperature drop $\Delta T$,
\begin{equation}
\Delta\rm T\le 1\ K,
\end{equation}
from virialization to the end of the simulation. This is a strict constraint used to categorize gas cloud evolution with nearly constant temperature after virialization. Meanwhile, the collapse condition in Eq.~\eqref{Jeans} is required to be fulfilled.
\item[Evaporation] In the presence of a very large UV flux, the ionization fraction at very high redshift becomes large enough to deplete the available neutral $\rm H$ for $\rm H_2$ formation. However, in the redshift range $200\le z<500$, the increased electron fraction temporarily enhances the formation of $\rm H^-$ through $\rm H+e^-\rightarrow H^-+\gamma$, thereby promoting $\rm H_2$ formation via $\rm H^-+\rm H\rightarrow H_2+e^-$. 
 
When this happens, we see that the temperature of the gas cloud drops precipitously well below the IGM temperature. While photoionization increases the free-electron fraction, enhancing Compton heating by the CMB, the increased ionization also promotes radiative cooling from bremsstrahlung. Together with more efficient adiabatic cooling due to the reduced mean molecular weight, these processes lead to a rapid temperature drop. When this happens, our simulations then abruptly stop  as \texttt{Grackle} hits its temperature floor. We categorize this case as a distinct outcome, which we refer to as evaporation, where the gas cloud becomes an overdense cold region in the early Universe. For the injection rates included in our work, we will not see this scenario occur unless self-shielding is removed, which would over-count the injected energy. We show results corresponding to this case in Appendix \ref{ss}, but caution that these correspond to an unphysical parametrization. 
\end{description}

\begin{figure}[!h]
\centering
\includegraphics[width=0.98\columnwidth]{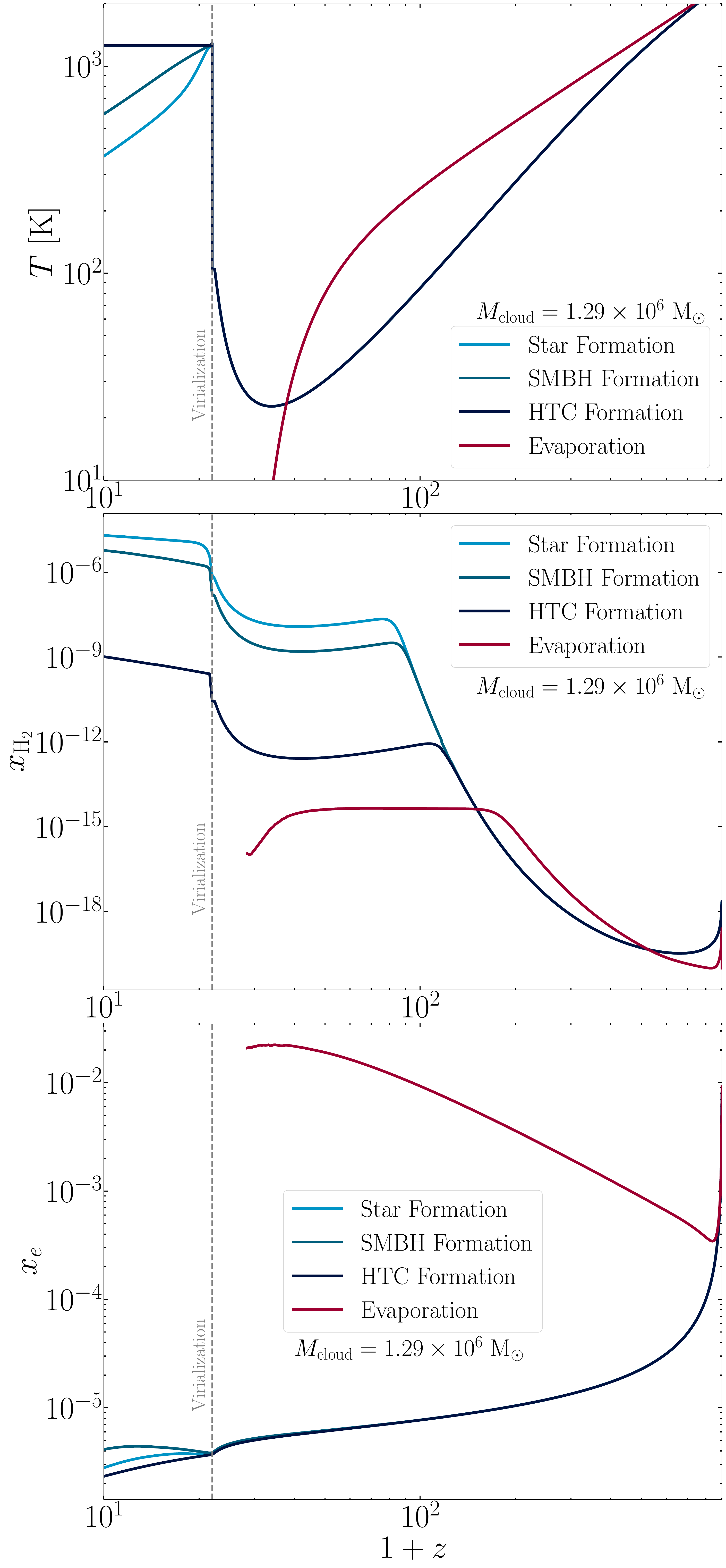}\\
\caption{Thermal and ionization histories of the gas cloud with $\rm M_{cloud}=1.29\times 10^6\ \rm M_\odot$  showing four distinct final stages with different DM decay parameters. For star formation, SMBH formation and evaporation, $m_\chi=23.55\ \rm eV$, $\tau=7.74\times10^{30}, 10^{30},  10^{22}\ \rm s$ respectively, for HTC formation, $m_{\chi}=11.43\ \rm eV$ and $\tau=2.78\times 10^{26}\ \rm s$. {\it Top:} Thermal histories of the gas cloud evolution. {\it Middle:} Fraction of $\rm H_2$ during the gas cloud evolution. {\it Bottom:} Free electron fraction in the gas cloud as a function of redshift. \label{fig:curves}}

\end{figure}
\section{Evolution of a single gas cloud}\label{results}
In this section, we present the thermal, ionization, and $\rm H_2$ evolution of gas clouds for the different outcomes outlined above, drawing maps in parameter space showing where these behaviours are expected to show up under the assumptions we have made. We then investigate the effects of DM UV injection on the formation of the first stars. 

\subsection{Behaviour of a single cloud}
In Fig.~\ref{fig:curves}, we illustrate the thermal and ionization histories, as well as the $\rm H_2$ fraction, during gas cloud evolution in the Star, SMBH seed, HTC formation, and evaporation scenarios for a $1.29\times 10^6,\rm M_\odot$ gas cloud and different choices of DM parameters. In the standard scenario without exotic UV injection, the only source of photons external to the chemical network is the CMB, whose flux rapidly becomes negligible with redshift, allowing for efficient H$_{2}$ assembly. Therefore, the evolution results in star formation due to the sufficiently rapid cooling of the gas cloud after virialization.

Increasing the UV flux allows for SMBH formation. This scenario shows similar behavior to the star-formation case until right after virialization. In both cases, the temperature of the gas cloud initially decreases due to adiabatic expansion, then increases as the gas cloud begins to collapse until it reaches the virial temperature $T_{\rm vir}\sim 1100\rm K$. The free electron fraction $x_e$ in both cases also closely follows the IGM value before virialization. The fractions of $\rm H_2$ in the gas cloud begin to differ at higher redshift around $z\sim 100$; however, since the free electron fraction at this time is still relatively small and energy release from adiabatic expansion still dominates, molecular cooling effects are negligible for the thermal history in both cases. After the gas clouds reach virialization, the density $\rho_{\rm gas}$ remains equal to $\rho_{\rm vir}$, fixing the internal energy. Meanwhile, the different $\rm H_2$ fractions lead to different cooling efficiencies, resulting in different thermal evolution.

With more $\rm H_2$ as coolant, the temperature of the gas cloud drops by $\sim275\rm K$ between virialization and $z=15$, which may trigger uneven cooling inside the gas cloud and eventually lead to fragmentation, contributing to the formation of the first stars. For DM with the same mass but shorter lifetime or larger cross section, the intensity of the external photon flux from DM decay/annihilation is stronger, resulting in more extreme suppression of $\rm H_2$, as shown in the middle panel of Fig.~\ref{fig:curves}, indicating insufficient coolant in the gas cloud and further leading to a more uniform cooling effect, thus ensuring SMBH seed formation. In both cases, there is a slight difference in the free electron fraction after virialization, as expected, due to the different photon fluxes.

If the suppression of $\rm H_2$ formation is more extreme, e.g., due to a more intense photon flux in the LW range, the fraction of $\rm H_2$ can be more strongly suppressed at high redshift, as shown by the HTC formation curve in the middle panel of Fig.~\ref{fig:curves}. At $z\sim 100$, the $x_{\rm H_2}$ is approximately four orders of magnitude lower than in the star and SMBH formation cases. This causes a deficiency of coolant during the evolution. After virialization, the temperature of the gas cloud remains almost invariant, while $x_{\rm H_2}$ is much lower than in the star or SMBH formation cases. We treat this as a separate case from SMBH formation due to the absence of effective cooling after virialization. However, the temperature of the gas cloud may eventually decrease due to particle escape, which leads to mass loss of the gas cloud and may induce delayed formation of a lighter SMBH seed. 

\begin{figure}[!htb]
\centering
\includegraphics[width=0.94\columnwidth]{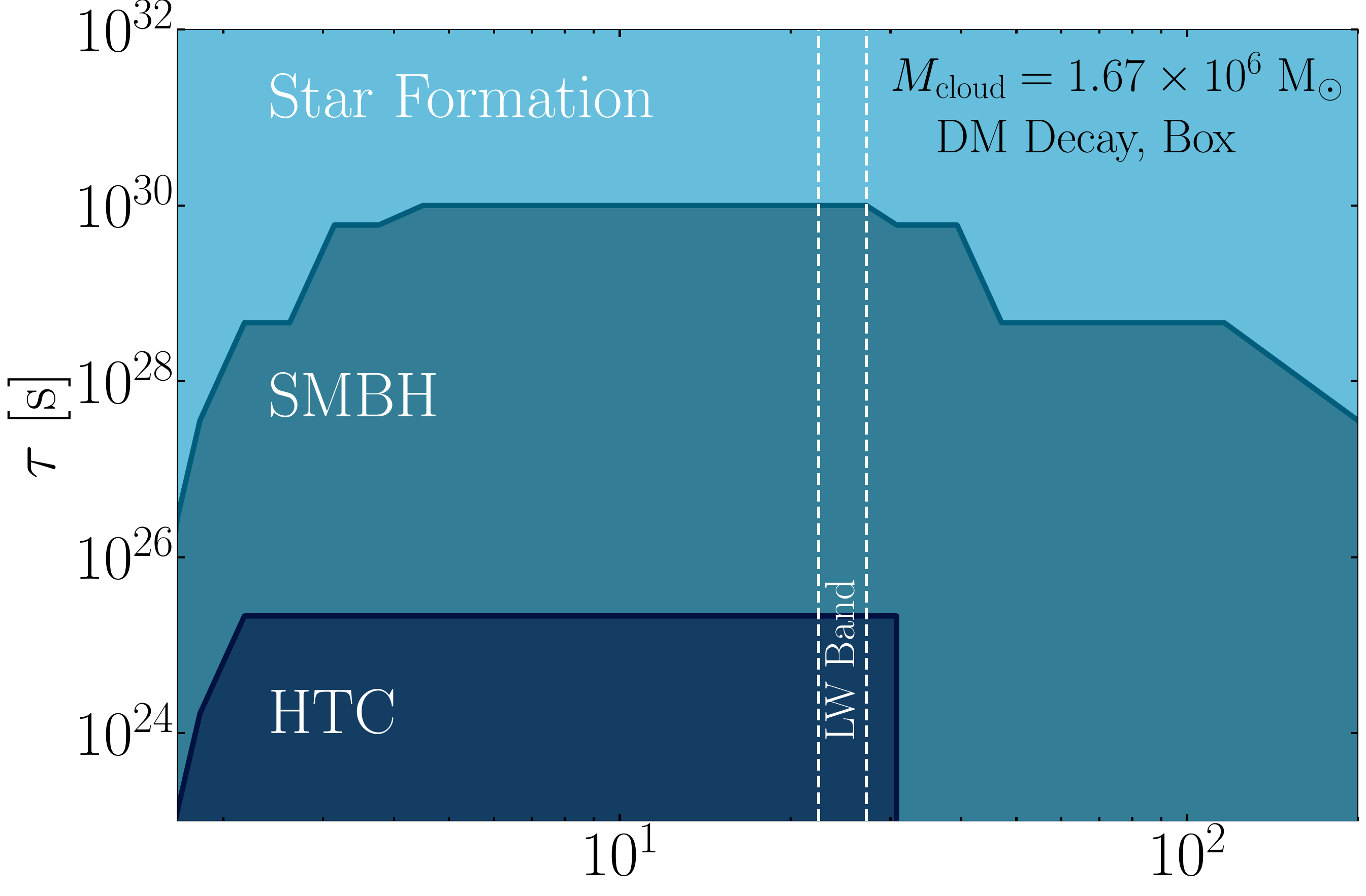} \quad
\includegraphics[width=0.94\columnwidth]{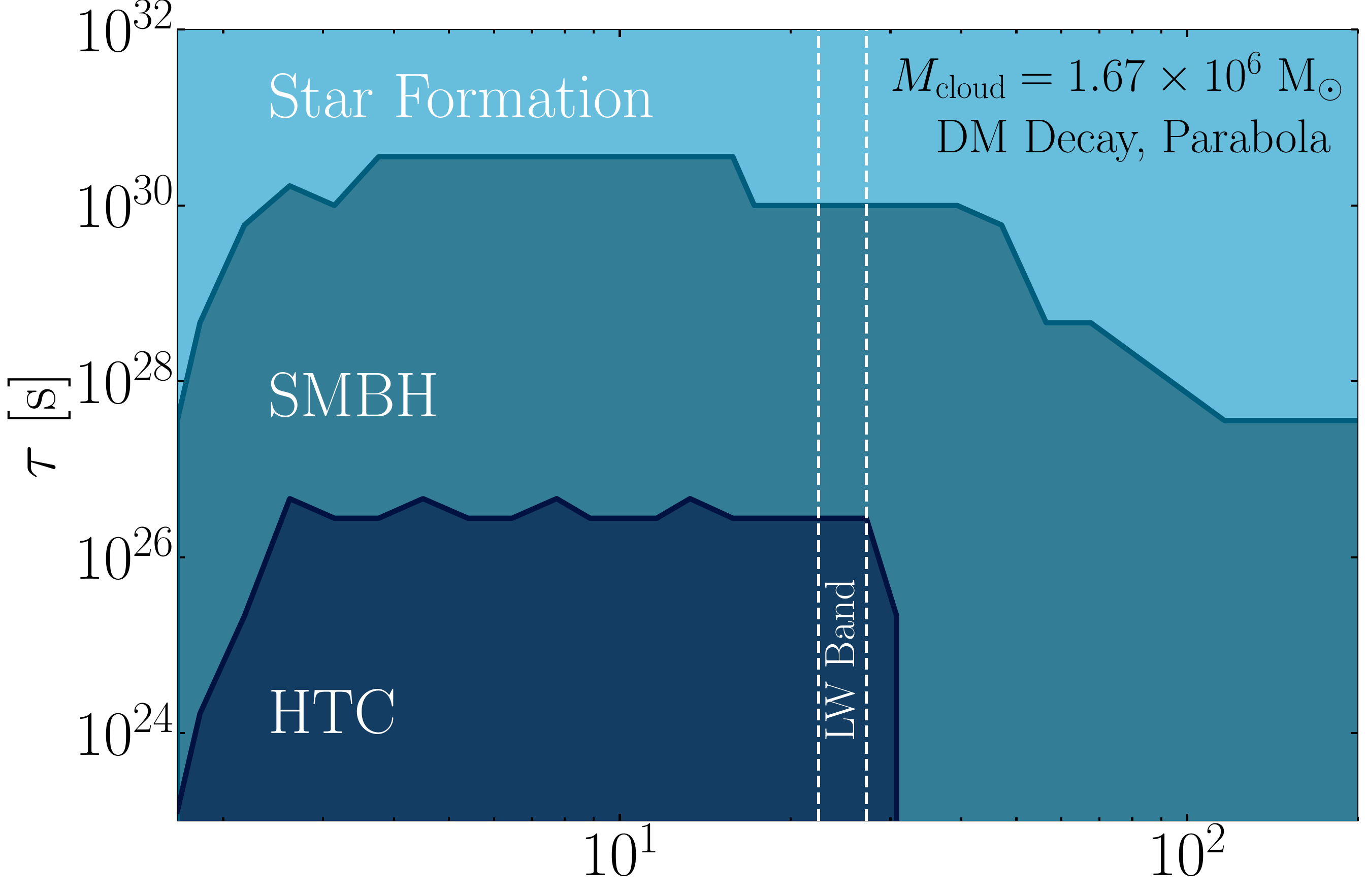} \quad
\includegraphics[width=0.99\columnwidth]{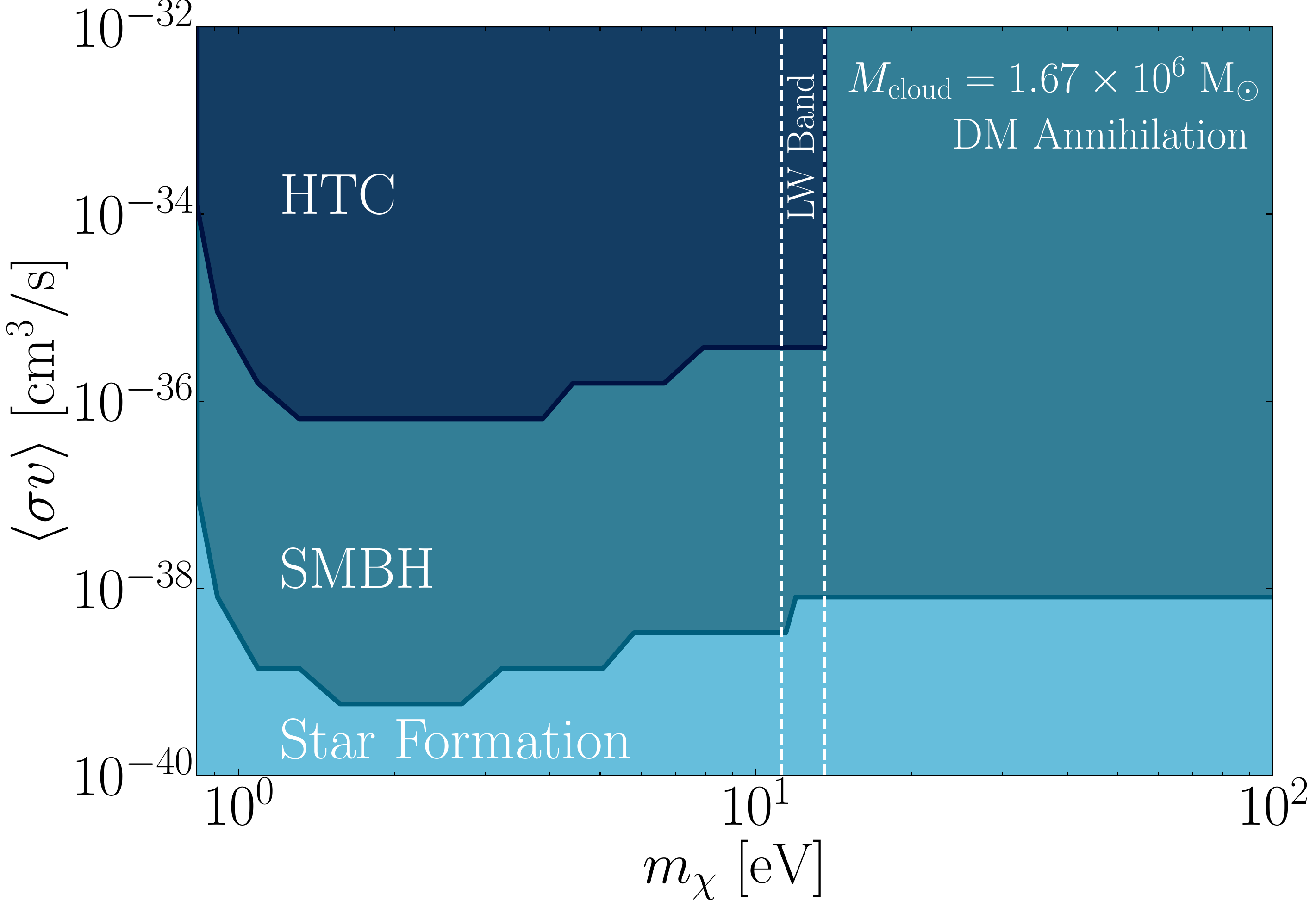} 
\caption{Phase diagrams of the gas cloud evolution as the parameter space of DM. White dashed lines give the mass range of DM that produces photons with energy in the LW band. The mass of the gas cloud is $1.67\times 10^6\ \rm M_\odot$. {\it Top:} DM decay with a box spectrum. {\it Middle:} DM decay with a parabola-shaped photon spectrum. {\it Bottom} Non-thermal Majorana DM annihilation. \label{fig:phasediagram}}

\end{figure}

Next, we scan the parameter space of both decaying and annihilating DM with gas cloud masses in $10^6\le M_{\rm gas}\le 10^7\ \rm M_\odot$: we illustrate results for decaying DM with box and parabola spectra, and annihilating DM with a parabola spectrum. In all cases, we will set $f_{dm} = f_\gamma = 1$. For decays, we examine the mass range $1.52\le m_{\chi}\le 200\ \rm eV$, and lifetimes $10^{22}\le\tau\le 10^{32}\ \rm s$; for annihilating DM we scan
$0.76\le m_\chi\le 100\ \rm eV $ and cross sections  $10^{-40}\le\langle\sigma v\rangle\le10^{-32}\ \rm cm^3/s$. 

Fig.~\ref{fig:phasediagram} shows the outcomes for a $1.67\times10^6\ \rm M_\odot$ gas cloud, including atomic and molecular self-shielding described in Sec.~\ref{sec:selfshielding}. The three regions that we see indicate the three possible outcomes: 1) in light blue, the unchanged history at large DM lifetimes/low cross sections, labeled as ``Star Formation'', 2) in intermediate shade, the possible formation of SMBH seeds due to suppressed H$_2$, and 3) in navy blue, the HTC outcome, due to excessive H$_2$ destruction, preventing any meaningful cooling. In all three figures, the white dashed line represents the range of masses that line up exactly with LW band edges. This is for visual reference, as any injection from masses above threshold will produce at least some photons in the band. In all figures, the lower edge of the mass axis corresponds to the photodetachment threshold of $m_\chi = 2\cdot 0.76$ eV for decay and $0.76$ eV for annihilation.

In the top panel of Fig.~\ref{fig:phasediagram}, we show the DM decay with a box spectrum. As the DM mass passes the photodetachment threshold, the lifetime at which star formation transitions to SMBH formation rises sharply to a plateau, where most of the energy is injected above threshold.

As $m_\chi$ increases past the LW band, the required decay rate increases again. Although the box-shaped spectrum provides a flat photon distribution over $0\le E_\gamma\le m_\chi/2$, the photon flux intensity decreases with increasing DM mass at fixed lifetime. Therefore, shorter lifetimes are required for SMBH formation, as more frequent DM decays produce sufficient photons to suppress $\rm H_2$ formation and maintain a moderate cooling rate. Within the window $1.66\le m_\chi\le30.72\ \rm eV$, DM with lifetimes $\tau\lesssim3\times10^{25}\ \rm s$ lead to enough $H_2$ suppression that collapse stalls completely, yielding the HTC scenario instead.

For DM with $m_\chi>27.2\ \rm eV$, the decay can produce photons above the H photoionization threshold, but their effects are limited by self-shielding from the high atomic-hydrogen number density, such that only a small fraction contributes to photoionization, meaning the extra $e^-$ population is not sufficient to significantly affect the outcome except HTC formation.

The middle panel of Fig.~\ref{fig:phasediagram} shows the equivalent outcome when decaying DM injects a parabola-shaped spectrum. Both nonstandard scenarios are triggered at longer lifetimes, thanks to extra photodetachment from photons in the central parts of the spectrum.

The bottom panel of Fig.~\ref{fig:phasediagram}  shows the phase diagram for gas cloud evolution with non-thermal Majorana DM annihilation as the source of the external photon flux. We observe what is effectively a mirrored version of the decay scenario, as annihilation increases as $\langle \sigma v \rangle$. The sensitivity of the cloud to annihilation peaks right above the photodetachment threshold, and then diminishes as higher-energy photons are injected, plateauing at higher masses since the energy injection rate in and below the LW band depends very weakly on the DM particle mass.

\subsection{Altered star formation history}

\begin{figure}[!htb]
\centering
\includegraphics[width=0.98\columnwidth]{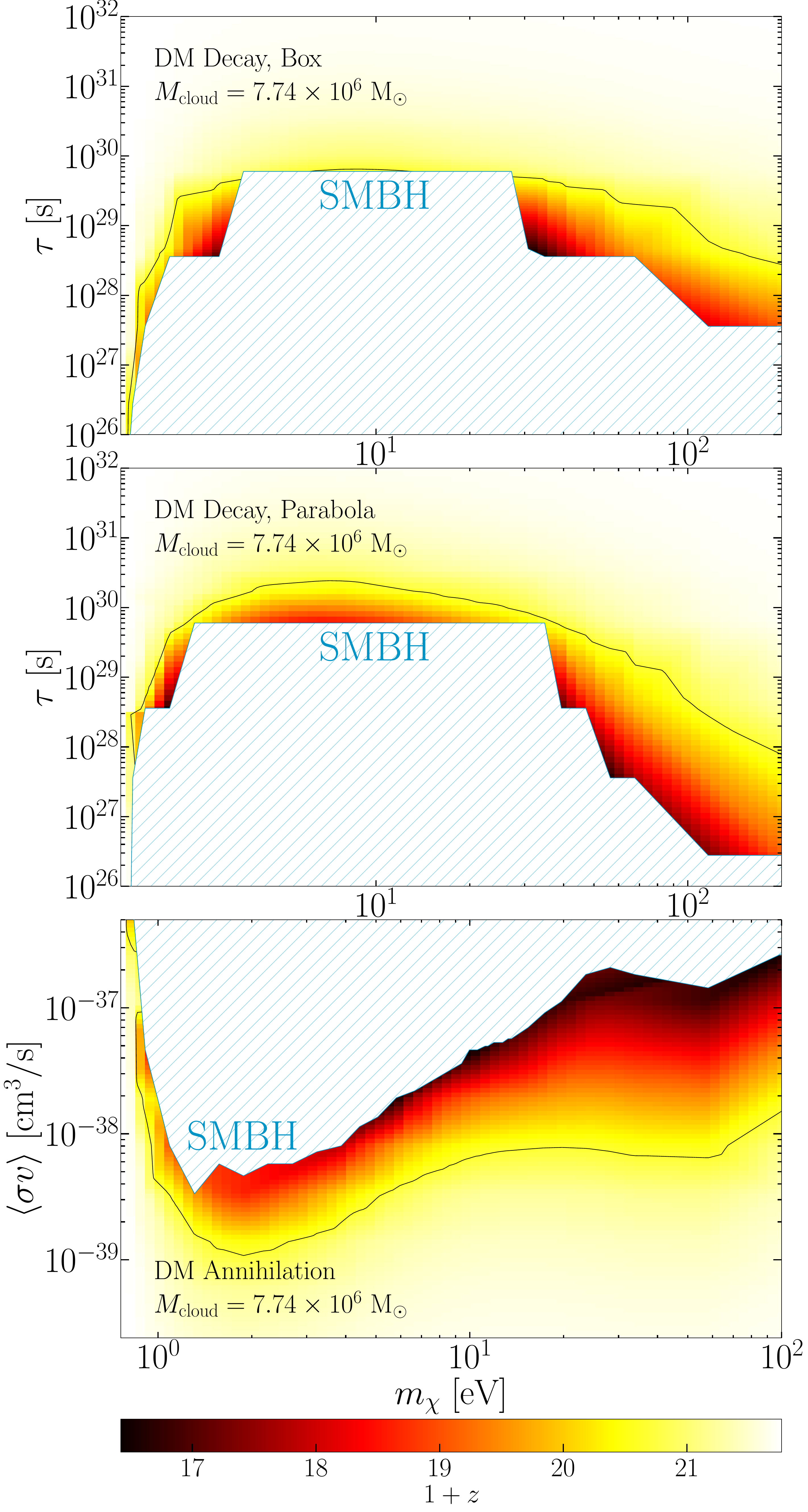}
\caption{Star formation time in our gas cloud evolution models as a function of the dark matter mass $m_\chi$ and lifetime (top two plots) or annihilation cross section (bottom plot). Black curves represent the boundaries of contours which star formation not later than $5\%$ of star formation time of SM (no DM) case. {\it Top:} DM decay with a box spectrum. {\it Middle:} DM decay with a parabola-shaped photon spectrum. {\it Bottom} Non-thermal Majorana DM annihilation.\label{fig:starformation}}

\end{figure}

Next, since the star-formation criterion is relatively well established, we determine the approximate star-formation redshift $z_{\rm SF}$ from $T(\eta z_{\rm SF})=\eta T(z_{\rm SF})$, with $\eta=0.75$. We then compare the star-formation time with the Standard Model (SM) case, in which the gas cloud evolves without external photon injection from DM. The resulting star-formation times are shown in Fig.~\ref{fig:starformation}, where hatched regions indicate no star formation. The black contour indicates the barrier below which star formation occurs at a redshift $5\%$ later than in the standard case. 
The top panel of Fig.~\ref{fig:starformation} illustrates the star-formation redshift for a $7.74\times10^6\ \rm M_\odot$ gas cloud exposed to photon injection from decaying DM with a box-shaped spectrum. The delay in star formation occurs starting at the photodetachment threshold $m_\chi\gtrsim1.52\ \rm eV$. The behavior follows the SMBH contours from Fig. \ref{fig:phasediagram}, with more delayed star formation as the DM lifetime becomes lower, with the peak occurring above the photodetachment threshold, near the photoionization energy.

The middle panel of Fig.~\ref{fig:starformation} shows the same results, but with a parabola-shaped spectrum. Again, the behaviour qualitatively mirrors the box spectrum, but with sensitivity to slightly higher lifetimes thanks to the spectral peak at intermediate energies near $m_\chi/2$. 

The bottom panel of Fig.~\ref{fig:starformation} shows the star-formation redshift for a $7.74\times10^6\ \rm M_\odot$ gas cloud exposed to photon injection from annihilating DM. Again, increased energy injection from larger cross section leads to inefficient cooling, delaying star formation until the SMBH collapse threshold is reached. The strongest sensitivity is for DM with a mass of a few eV, which maximises the number of photons injected. The strongest effect is clearly from photodetachment, rather than photodissociation in the LW band. 

These collapse outcomes and star formation redshift results include the full self-shielding parametrisation of Sec. \ref{sec:selfshielding}. These become qualitatively very different when self-shielding is removed, as the extra ionizing radiation for large DM masses leads to extra cooling, and thus either rapid star formation or evaporation.

\begin{figure}[!htb]
\centering
\includegraphics[width=0.98\columnwidth]{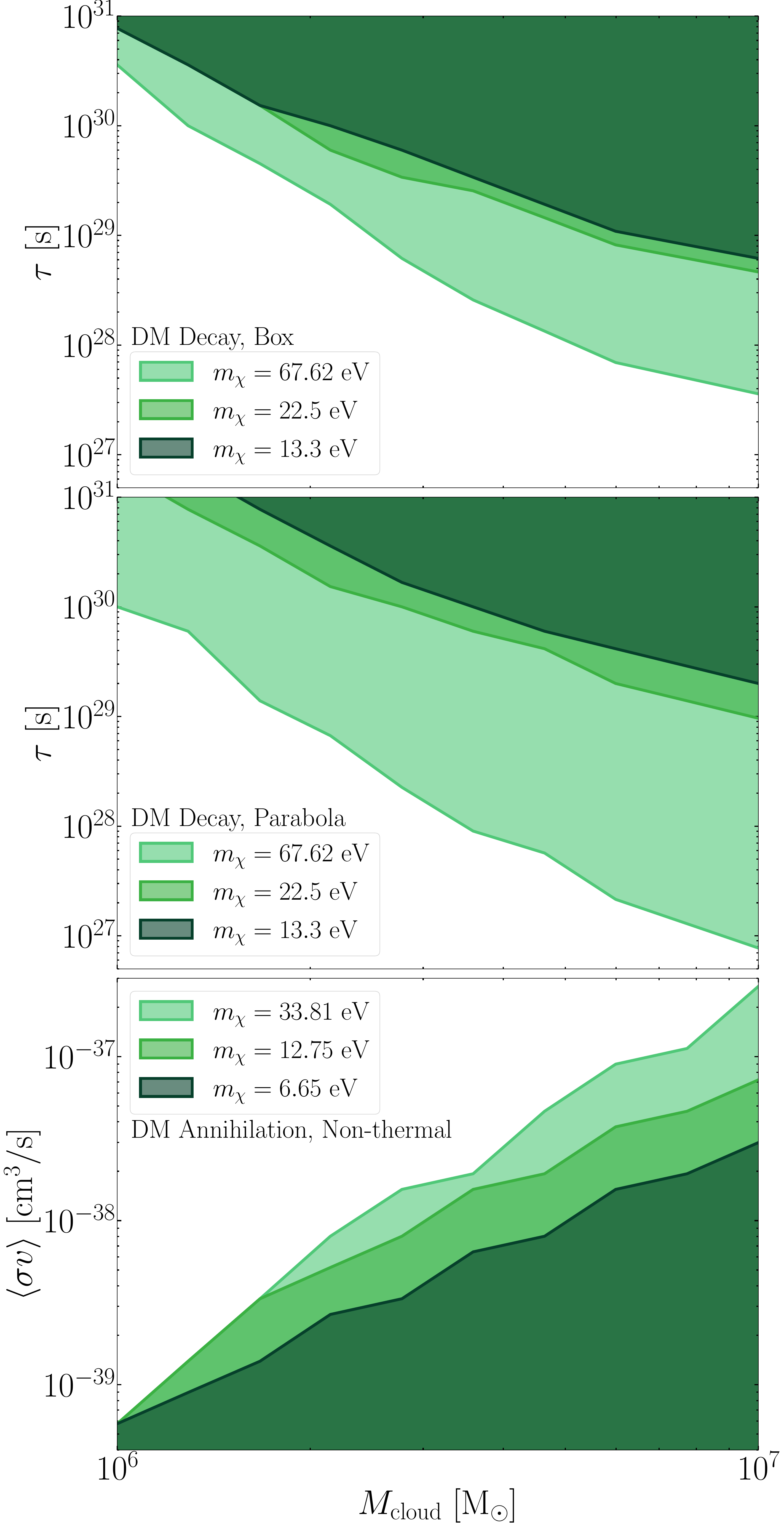}
\caption{Critical cloud mass curves of the star formation with specific DM mass. {\it Top}: DM decay with a box spectrum. {\it Middle:} DM decay with a parabola-shaped photon spectrum. {\it Bottom} Non-thermal Majorana DM annihilation. \label{fig:cloudmass}}

\end{figure}

Finally, we show the effect of varying the halo mass for fixed DM mass. This results in star formation ``critical curves'', which we show in  Fig.~\ref{fig:cloudmass}, for decay to a box spectrum (top), parabola spectrum (middle), and annihilation (bottom).  In each panel, the lines indicate the threshold below which star formation is fully suppressed---this can also be interpreted as a threshold for DCBH formation. Lower halo masses require less energy injection, and therefore are more sensitive to extra UV injection from DM annihilation. 

Reading these as thresholds for BH direct collapse, this means that for a given decay rate or annihilation cross section, there will be a ceiling mass above which DCBH from exotic energy injection no longer works. This should in turn lead to a maximum seed size, and thus a distinctive high-mass cutoff in the DCBH mass function.

We end by commenting on differences between these results and those presented in Ref. \cite{Qin:2023kkk}. In that work, the authors examined the star formation time as affected by the dominant external flux of photons from the decay of heavier DM in the 10 keV--TeV mass range. In this case, the cascade from higher energy injection products produces a much higher flux of ionizing radiation and e$^\pm$. This can lead to free electron fractions that are significantly higher than what is obtained here, catalyzing the formation of extra H$_2$ coolant. In competition with the extra heating due to these products, this can lead both to delayed star formation (similar to what we see here), as well as accelerated star formation thanks to the additional H$_2$. 

\section{SMBH seed mass function}
\label{sec:BHMF}
Given these results, we turn to the question of what would be required for DM-catalyzed black hole seeds to explain the observed abundance of high-redshift SMBHs.

After formation, an SMBH seed grows into a SMBH at lower redshift by accreting gas at a rate that is proportional to its bolometric luminosity. This accretion process can be described as a relation between the mass of the seed and the SMBH, as well as the accretion time~\cite{Bhattacharya:2025dgx}
\begin{equation}
    M_{\rm BH}^{\rm obs}=M_{\rm seed}\exp\left[\frac{\eta_{\rm acc}\left(1-\epsilon_r\right)\left(t_{\rm obs}-t_{\rm seed}\right)}{t_{sal}}\right],
    \label{eq:singleacc}
\end{equation}
where $t_{\rm sal}\approx45.1\left(\epsilon_r/0.1\right)\ \rm Myr$ is the Salpeter time,  the radiative efficiency is set to $\epsilon_r=0.1$, $\eta_{\rm acc}$ is the accretion efficiency, where $\eta_{\rm acc}=1$, $\eta_{\rm acc}<1$ and $\eta_{\rm acc}>1$ correspond to Eddington accretion, sub-Eddington accretion and super-Eddington accretion, respectively. This parametrization has been widely used in previous works and provided plausible fits to simulation results~\cite{Bramberger:2015kua, Bhattacharya:2025dgx}. However, its accuracy suffers when the difference $t_{obs}-t_{seed}$ is large, due to the exponential growth of both accretion rate and by extension its associated uncertainty.

Ref~\cite{Willott:2010yu} further showed that the luminosity of early seeds appears to indicate near-Eddington accretion, but that it gradually slows down with redshift. Therefore, we utilize a modified time-dependent accretion efficiency 
\begin{equation}
    \eta_{\rm acc}(t_{\rm obs})=\eta_{\rm floor}+\frac{1-\eta_{\rm floor}}{1+\exp\left[\left(t_{\rm obs}-t_{\rm c}\right)/\tau_{\rm acc}\right]},
    \label{eq:acceta}
\end{equation}
where $t_{\rm c}$ is the time of suppression onset and $\tau_{\rm acc}$ is the transition width.
If we then start with seeds in mass range  $2.5\times10^5-10^7\ \rm M_\odot$  at redshift $20$ and evolve their accretion history forward, we end with a range of plausible SMBH masses in the redshift range of observed quasar candidates. The upper boundary of this range corresponds to the largest seed we are able to produce given Fig. \ref{fig:cloudmass}: any larger, for a given set of DM parameters, and cooling becomes too rapid. At the lower end, the gas cloud reaches its minimal mass to form an SMBH seed. The parametrizaion also assumes that the full gas cloud's mass collapses to form the seed. In reality, the BH seed will be a fraction of this mass. This will ultimately be equivalent to a logarithmic change in the transition time, so we set this fraction to one for simplicity.

\begin{figure}[!htb]
\centering
\includegraphics[width=0.98\columnwidth]{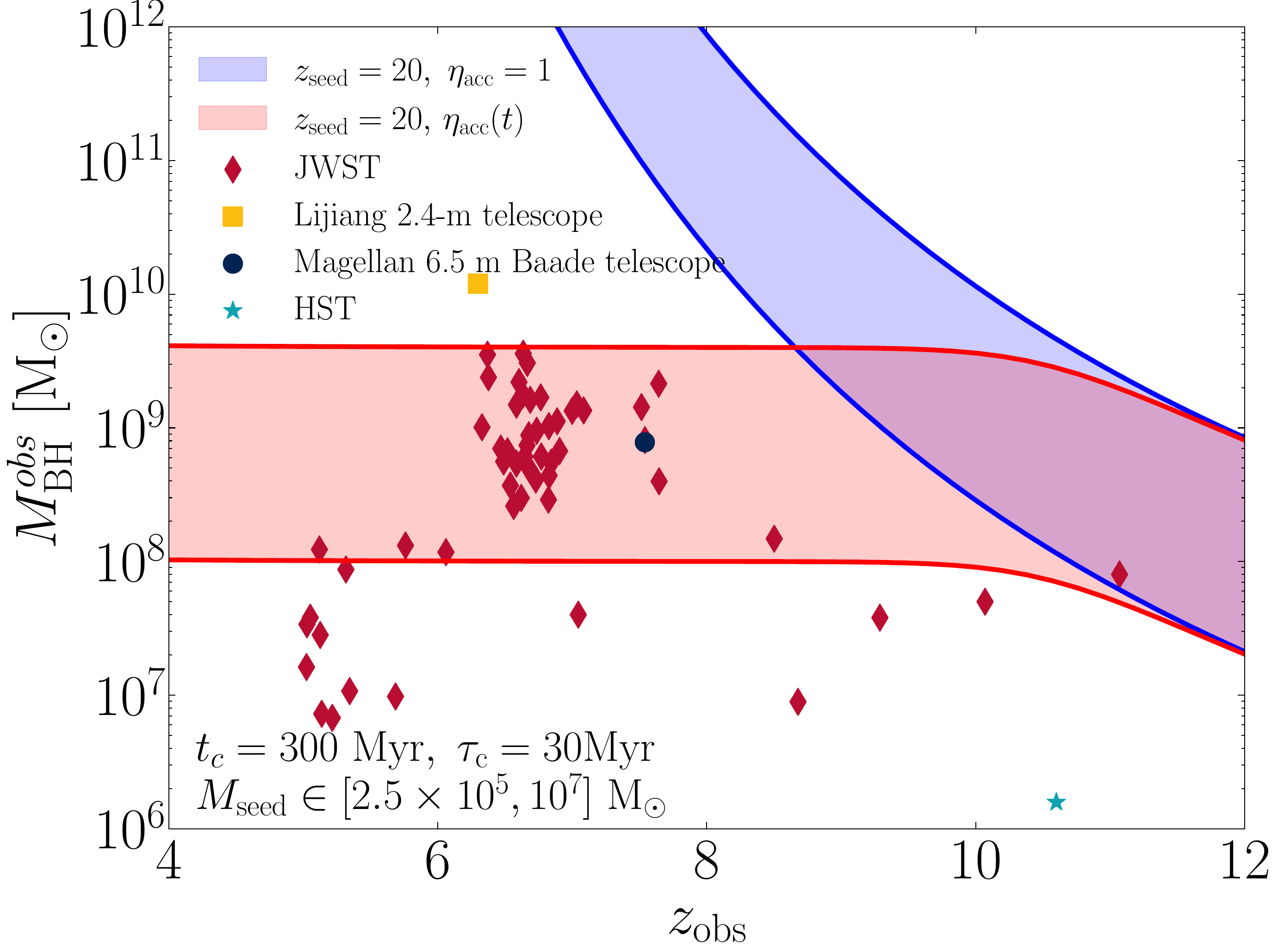}
\caption{Accretion history of the SMBH seeds with mass between $2.5\times10^5-10^7\ \rm M_\odot$. The blue band corresponds to the accretion history of SMBH seeds which reach virialization at $z=20$ with Eddington accretion efficiency $\eta_{\rm acc}=1$ during the whole accretion period, therefore the mass of observational SMBH increases exponentially with the redshift and the mass of SMBH becomes too large at redshift $z_{\rm obs}\sim8$. Red band represents the accretion of SMBH seeds which formed at $z=20$ with a time-dependent accretion efficiency, where the transition onset $t_c=300\ \rm Myr$ and transition width $\tau_c=30\ \rm Myr$ are chosen to emcompass the largest observed SMBHs. For comparison, we show observational data from telescopes: Lijiang 2.4-m telescope~\cite{Wu:2015phw}, Magellan 6.5m Baade telescope, Hubble Space Telescope~\cite{Oesch:2016dys} and JWST~\cite{Banados:2017unc, Oesch:2016dys, napolitano2025dual, Taylor:2025ubx, Bogdan:2023ilu, larson2023ceers, goulding2023uncover, furtak2023jwst, lin2026bridging, Yang:2021imt}.\label{fig:accretion}}

\end{figure}

Fig.~\ref{fig:accretion} shows this accretion history, plotted on top of observed quasars between redshifts 5 and 11. The upward-trending purple band shows unfettered Eddington accretion, while the pink band shows the effects of suppression \eqref{eq:acceta}. Here, we have picked $t_c = 300$ Myr so that the top of the band encompasses most of the observed massive quasars. We choose $\tau_{\rm acc} = 30$ Myr, though in practice this only serves to smooth the turnoff, with very little consequence on observables. 

Next, we turn to a toy model of the full SMBH population. Given a map between the SMBH seed mass at redshift $z_{\rm seed}$ and its associated dark matter halo mass, we can estimate the number density of SMBH seeds, as it will be proportional to the halo mass function (HMF) at that redshift. 
We then evolve this population forward in time according to the prescription above, to give a redshift-dependent BH mass function (BHMF).   In reality, two seeds with the same mass will not necessarily share the same accretion history, which we account for by defining a stochastic accretion process. We draw the suppression time $t_c$ from a Gaussian distribution
\begin{equation}
    p_c(t_c)= \frac{1}{\sqrt{2\pi}\sigma_{t_c}}\exp\left[-\dfrac{(t_c-\mu_{t_c})^2}
{2\sigma_{t_c}^2}\right]\,.
\end{equation}
The midpoint $\mu_t$ is a parameter that we will float to match observed SMBH data, along with a normalization $f_{\rm SMBH}$ that parametrizes the fraction of host halos that successfully seed a black hole.\footnote{In practice we of course do not allow $t_c$ to be less than zero, and adjust the normalization accordingly. This makes very little difference for the hierarchy of $\sigma_{t_c}$ and $\mu_t$ that we consider here.} The likelihood is insensitive to $\sigma_{t_c}$ over the range considered here, so we fix it to 20 Myr.

We use the HMF parametrized in Ref.~\cite{lopez2013constraints}. 
Then with the stochastic accretion process, we map the HMF to the BHMF by including a probability distribution function (PDF):
\begin{equation}
\begin{aligned}
\frac{dn_{\rm SMBH}}{dM_{\rm SMBH}}
={}&\frac{f_{\rm SMBH}}{f_b}
\int_{M_{\rm min}}^{M_{\rm max}}
dM_{\rm seed}\,
\left.
\frac{dn_{\rm halo}}{dM_{\rm halo}}
\right|_{M_{\rm halo}=M_{\rm seed}/f_b}
\\[4pt]
&\times K(M_{\rm SMBH}\mid M_{\rm seed};\Delta t),
\end{aligned}
\end{equation}
where $\Delta t$ is time interval between $z_{\rm obs}$ and the seed formation redshift $z_{\rm seed}$, $f_{\rm SMBH}$ is the fraction of halos with $M_{\rm halo}=M_{\rm seed}/f_b$ that can evolve to SMBHs with mass $M_{\rm SMBH}$, $K(M_{\rm SMBH}\mid M_{\rm seed};\Delta t)$ represents the fraction of seeds that grow from $M_{\rm seed}$ to $M_{\rm SMBH}$  in the time interval $\Delta t$:
\begin{equation}
\label{eqs:kernel}
\begin{aligned}
K(M_{\rm SMBH}\mid M_{\rm seed};\Delta t)
=\int_0^\infty dt_c p_c(t_c)\,
\\
\delta_{\rm D}\!\left[
M_{\rm SMBH}-F(M_{\rm seed},t_c;\Delta t)
\right],
\end{aligned}
\end{equation}
where the SMBH seed accretion process follows directly from Eq. \eqref{eq:singleacc}:
\begin{equation}
    F(M_{\rm seed},t_c;\Delta t)=M_{\rm seed}\exp\left[\frac{\left(1-\epsilon_r\right)}{t_{sal}}\int_0^{\Delta t}\eta_{\rm acc}(t, t_c)dt\right] \,.
\end{equation}
Since the mass accretion process is monotonic, Eq.~\eqref{eqs:kernel} can be expressed as
\begin{equation}
    K(M_{\rm SMBH}\mid M_{\rm seed};\Delta t)
=\left|\frac{\partial F}{\partial t_c}\right|_{t_c^*}^{-1}p_c(t_c^*),
\end{equation}
where the prefactor comes from the Jacobian transforming the Dirac delta in Eq. \eqref{eqs:kernel} to a function of time, and $t_c^*$ is the value of $t_c$ where $M_{\rm SMBH} = F(M_{\rm seed},t_c;\Delta t)$.
\begin{table}
\centering
\caption{Best-fit parameters and $1\sigma$ profile-likelihood intervals
($\Delta\chi^2 \le 1$), with $\tau_{t_c}$ fixed at $30$\,Myr.\label{tab:bhmffits}}

\begin{tabular}{lccccc}
\toprule
Dataset & $z_{\rm obs}$ & $\chi^2_{\rm min}$ & $\sigma_{\rm int}$ & $\mu_{tc}$ [Myr] & $\log_{10} f_{\rm SMBH}$ \\
\midrule
SDSS-M   & 6    & 1.28 & 0 & $455.5^{+4.2}_{-10.2}$ & $-9.59^{+0.07}_{-0.06}$ \\
SDSS-O   & 6    & 0.042 & 0 & $435.7^{+6.5}_{-10.2}$  & $-8.99^{+1.33}_{-0.49}$ \\
SDSS-DR7 & 4.75 & 3.12 & 0.614 & $424.1^{+3.8}_{-9.2}$ & $-8.16^{+0.17}_{-0.12}$ \\
JWST      & 5    & 0.0034 & 0 & $218.1^{+4.4}_{-3.2}$ & $-4.77^{+0.07}_{-0.06}$ \\
HSC       & 4    & 23.3 & 1.402 & $322.0^{+2.3}_{-4.2}$   & $-5.08^{+0.29}_{-0.24}$ \\
\midrule
Joint     & --  & 181.2 & 2.32 & $297.0^{+3.7}_{-1.0}$   & $-5.33^{+0.84}_{-0.88}$ \\
\bottomrule
\end{tabular}
\end{table}

We fit our model to observational black hole mass functions (BHMFs) inferred from quasar samples observed by the Sloan Digital Sky Survey (SDSS)~\cite{Wu:2022njo, Shen:2011ge}, JWST~\cite{Matthee:2023utn} and the Hyper Suprime-Cam Subaru Strategic Program (HSC)~\cite{he2024black}.  These datasets probe different redshift and black hole mass ranges and are subject to different selection effects and systematic uncertainties. The SDSS-based BHMF datasets are derived from selected quasars with FWHM-based virial mass estimates, while the JWST constraints probe broad-line faint active galactic nuclei (AGNs) at high redshift by including JWST/NIRCam imaging and wide field slitless spectroscopy from the EIGER and FRESCO surveys. The HSC-SSP dataset analyzed less-luminous quasars data and extends the BHMF constraints towards lower black hole masses at $z\sim4$. The inferred BHMFs from each of these datasets may differ in their normalizations due to redshift evolution, differences in quasar selection and completeness corrections, and systematic uncertainties in black hole mass estimates. We therefore fit each dataset separately at its corresponding redshift, in addition to performing a joint fit of all datasets with observations at redshift $z_{\rm obs}= 6$.

We fit our toy model to these datasets by maximizing the likelihood constructed from the observational BHMFs, using a split-normal distribution to account for different upper and lower error bars. We find that the transition width, representing how long the transition from Eddington accretion takes, has very little impact on the likelihood, so we fix $\tau_{t_c}=30\ \rm Myr$. Because of the inconsistency between datasets, our fits to SDSS-D7, HSC and the joint fit return large $\chi^2$ values. For these, we introduce a zero-mean Gaussian intrinsic scatter $\epsilon_{\rm int}$,
\begin{equation}
    \log_{10}\Phi_{\rm obs}=
\log_{10}\Phi_{\rm model}+\epsilon_{\rm int},
\end{equation}
where
\begin{equation}
    \Phi\equiv\frac{dn_{\rm SMBH}}{d\log M_{\rm SMBH}}.
\end{equation}
The standard deviation $\sigma_{\rm int}$ of this scatter is added in quadrature to the upper and lower observational uncertainties in $\log_{10}\Phi_{\rm model}$,
\begin{equation}
    \sigma_{\pm,\rm eff}=\sqrt{\sigma_{\pm,\rm obs}^{\,2}+\sigma_{\rm int}^{\,2}}
\end{equation}
to account for potential deviations beyond the measurement errors reported. We fit $\sigma_{\rm int}$ simultaneously with the other model parameters.

\begin{figure*}[!htb]
\centering
\includegraphics[width=0.8\textwidth]{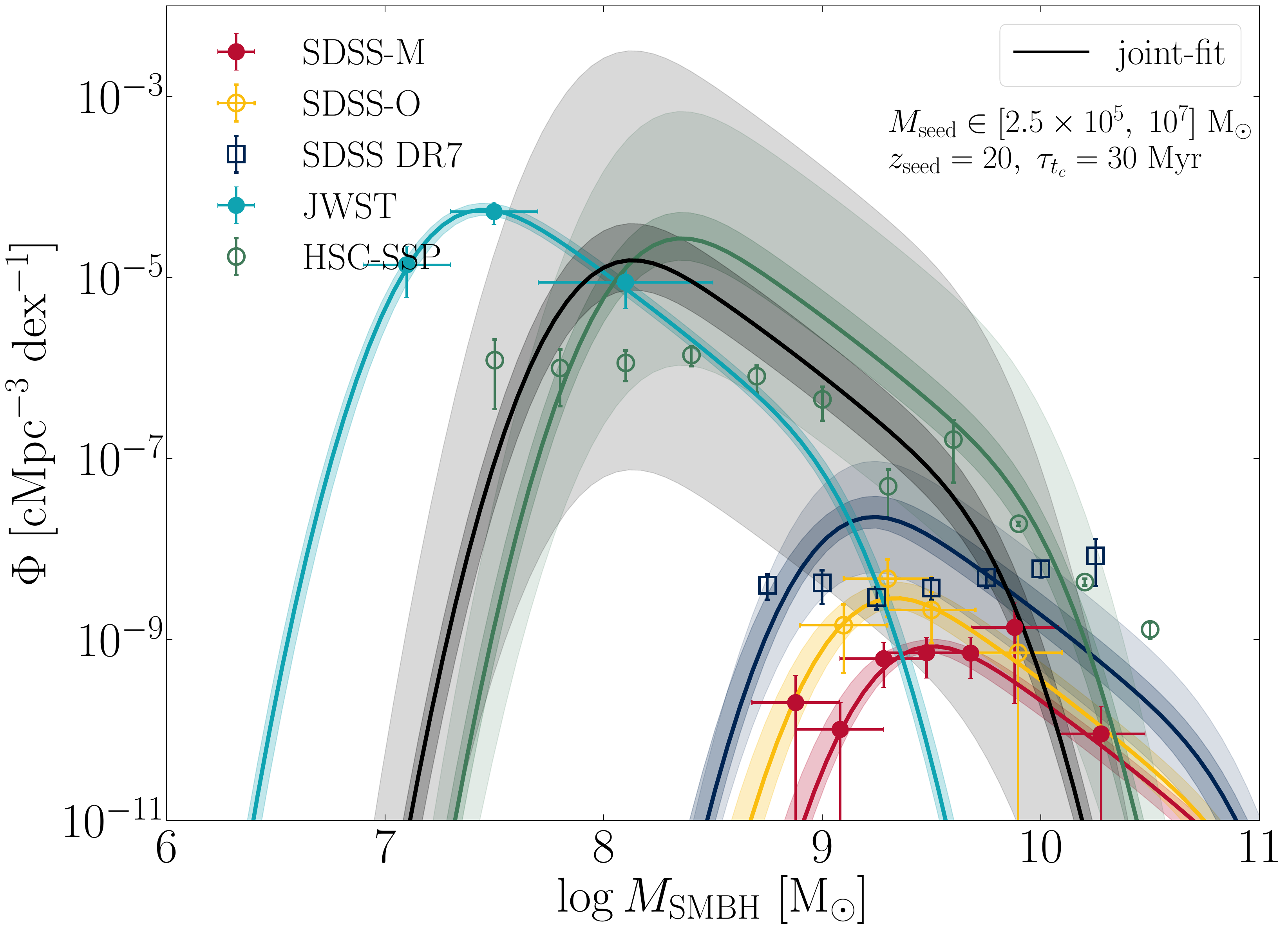}
\caption{
 Best-fit black hole mass function (BHMF) $\Phi = {dn_{\rm SMBH}}/{d\log M_{\rm SMBH}}$ curves with $1\sigma$ profile-likelihood intervals. Datasets from SDSS main (SDSS-M) and overlap (SDSS-O), SDSS DR7 quasar catalog, JWST/NIRCam and HSC-SSP. Black curve shows the joint best-fit of all datasets at redshift $z_{\rm obs}=6.$ Darker shading show the $1\sigma$-uncertainties of fitting, lighter shading show the error range with intrinsic-scatter $\sigma_{\rm int}$ for potential deviations of fitting.\label{fig:massfunc}}

\end{figure*}

We give the best-fit parameters $\mu_{t_c}$ and $f_{\rm SMBH}$ in Table.~\ref{tab:bhmffits} with $1\sigma$ errors. The resulting fitting curves with $1\sigma$ error bands and $\sigma_{\rm int}$ uncertainties are shown in Fig.~\ref{fig:massfunc}. 
As the figure shows, our model provides good fits to the high-redshift BHMF datasets (SDSS-M, SDSS-O, and JWST), whereas the fits to the low-redshift BHMF datasets are poorer. In SDSS-based datasets, both transit time expectation and $f_{\rm SMBH}$ fitting parameter are close, this may come from the similar quasar selection and completeness corrections. Shorter $\mu_{t_c}$ in JWST and HSC datasets indicate shorter Eddington accretion phase during seed growth, while higher $f_{\rm SMBH}$ values imply that larger fractions of the initial halos are required to form SMBH seeds to reproduce the observed BHMFs.

We also provide the fitting parameters for joint fitting by including all datasets. The resulting $\mu_{t_c}\sim 300\ \rm Myr$ shows shorter Eddington accretion period than heavier SMBHs with $M_{\rm SMBH}\sim 10^9-10^{10}\ \rm M_\odot$ in SDSS-based datasets. $f_{\rm SMBH}\sim10^{-5.33}$ implies that only a small fraction of halos are qualified to produce SMBH seeds in our model. This low fraction may reflect the combined requirements of large initial overdensities to seed these BHs and of a gas-rich environment at lower redshift. The joint fit provides a poor description of the datasets, which probe different redshifts and are subject to different selection effects and systematic uncertainties. 

The better agreement with the high-redshift datasets may be consistent with a scenario in which direct collapse contributes to the formation of the SMBHs in high-redshift quasars. However, this interpretation is not unique, since the observed BHMF itself has substantial systematic and statistical uncertainties from $1/V_{\max}$ estimation and our BHMF is obtained from a simple modification of accretion efficiency. Moreover, the SMBHs at the lower redshift may have undergone different evolutionary pathways, including substantial growth through mergers, which are not  accounted for in our model. Therefore, the best-fit parameters listed in Table~\ref{tab:bhmffits} only represent the parameter values required by our model to reproduce each observational BHMFs.  

These results represent the predictions from a one-zone model and a simplified few-parameter accretion model. Of course, a more complete answer would require tracing the evolution from initial overdensity fluctuations all the way to the redshift of SMBH observation, as well as to account for the merger history. Regardless, these provide an initial proof of concept along with rough parameters that indicate the efficiency and timescales required for DM to catalyze the population of observed SMBHs. The small fractions $f_{\rm SMBH}$ indicate that even if this scenario is rare or easily disrupted, the DM-induced direct collapse scenario may prove robust.

\section{Conclusion}\label{conclusion}
In this work, we have explored the possible consequences of a light dark species decaying or annihilating to photons on the collapse of supermassive black hole seeds at cosmic dawn. We constructed a metal-free primordial gas cloud model with external energy injection induced by DM in a DM halo and cosmological background, then presented its evolution from high redshift using a complete chemical network and its effects on the thermal history. 

Scanning the parameter space of decaying and annihilating DM, we identify three possible outcomes of primordial gas cloud evolution: (i) first star formation, driven by rapid $\rm H_2$ cooling after virialization; (ii) SMBH seed formation via direct collapse, driven by moderate cooling, which could account for the massive quasars observed at high redshift; and (iii) HTC formation, in which very weak cooling leaves the cloud temporarily hot, possibly leading to a lower-mass SMBH seed later after evaporative mass loss. DM-sourced photons in the LW band promote direct-collapse seed formation by photodissociating $\rm H_2$ and photodetaching $\rm H^-$. For annihilating DM with $0.76 \le m_\chi \le 100\ \rm eV$, cross sections $\gtrsim 10^{-38}\ \rm cm^3/s$ yield enough UV flux for SMBH seed or HTC formation; for decaying DM, lifetimes as low as $\sim 10^{30}\ \rm s$ suffice depending on the DM mass, with the parabola spectrum allowing a narrower viable region than the box spectrum, particularly at high $m_\chi$. Hydrogen self-shielding prevents an overestimate of high-energy flux that would otherwise spuriously prevent collapse. Using a simple model of accretion, and a map from the halo mass function at virialization onto a black hole mass function, we have shown that the observed population of high-redshift SMBHs can in principle be explained by these DM-catalyzed BHs, though more observations and more precise modeling will be necessary for a definitive answer.

DM under the consiered model can also delay ordinary star formation, potentially pushing it down into the reach of current 21\,cm experiments. Heavier clouds, which contain more $\rm H_2$, require larger DM photon fluxes (shorter lifetimes or larger cross sections) to form seeds directly, providing an upper limit to the reach of the DCBHs that can be produced in this way. Finally, while the one-zone model captures the early collapse, it cannot resolve post-virialization density evolution below the Jeans or filtering scale \cite{Gnedin:2000uj}; higher-dimensional hydrodynamical simulations would be required to fully understand this process.

The dark matter masses considered here also lie in an awkward region of parameter space. Such models would require nonthermal production mechanisms, or would need to be a subcomponent of the dark matter in order to evade a panoply of constraints, which is possible but somewhat baroque \cite{Friedlander:2022ovf}. We chose them simply because the energies map most easily to the relevant energies to directly affect the chemical network in the cooling primordial gas. To consider more standard WIMP/FIMP scenarios, we would need to accurately model the energy cascade down from GeV annihilation products to the UV. Though this has been done for energy injected in the homogeneous cosmological background \cite{Qin:2023kkk}, the intra-halo particle cascades depend on the local density and composition, as well as the redshift as inverse-Compton losses on the CMB remain important \cite{Schon2018,Hou:2024hcj}. This will be the subject of future work.

\acknowledgements
We thank Robert Brandenberger, Ben Lehman, Boyuan Liu, Katie Mack, Wenzer Qin, Tracy Slatyer, Yitian Sun and James Taylor for useful discussions. This work was supported by the Arthur B. McDonald Canadian Astroparticle Physics Research Institute, NSERC, the Ontario Ministry of Universities and Colleges, the Canada Foundation for Innovation, and the Queen’s Centre for Advanced Computing. This research was supported in part by Perimeter Institute for Theoretical Physics. Research at Perimeter Institute is supported by the Government of Canada through the Department of Innovation, Science and Economic Development and by the Province of Ontario through the Ministry of Research, Innovation and Science. 

\appendix

 \section{Phase Diagrams and Star Formation Time with no Self-Shielding Case}\label{ss}
In this section, we present the phase diagrams of gas cloud evolution across the DM parameter space, as well as the star formation time, neglecting the effects of self-shielding.
\begin{figure}[!htb]
\label{fig:phase_diagram_noss}
\centering
\includegraphics[width=0.94\columnwidth]{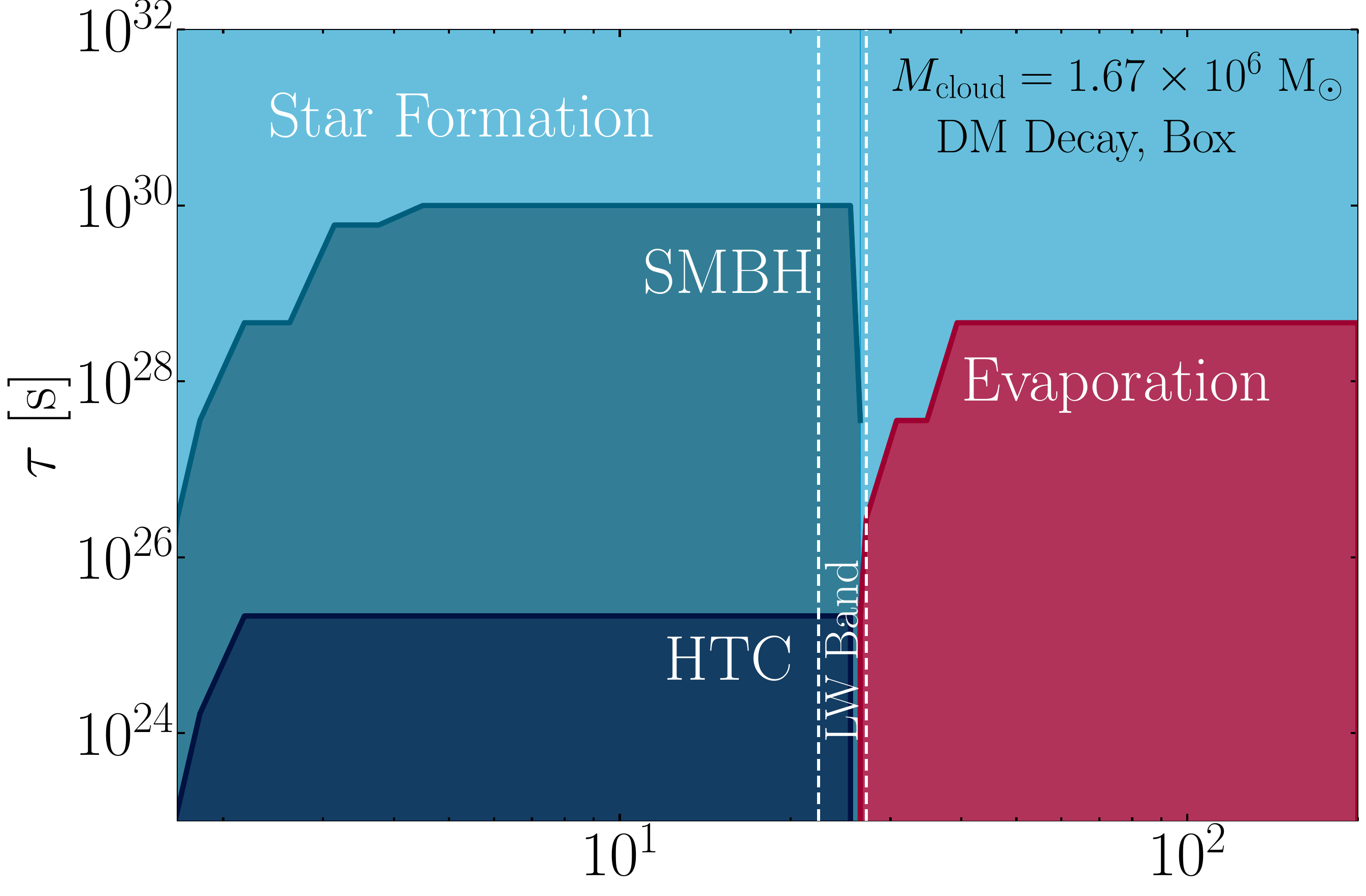} \\ 
\includegraphics[width=0.94\columnwidth]{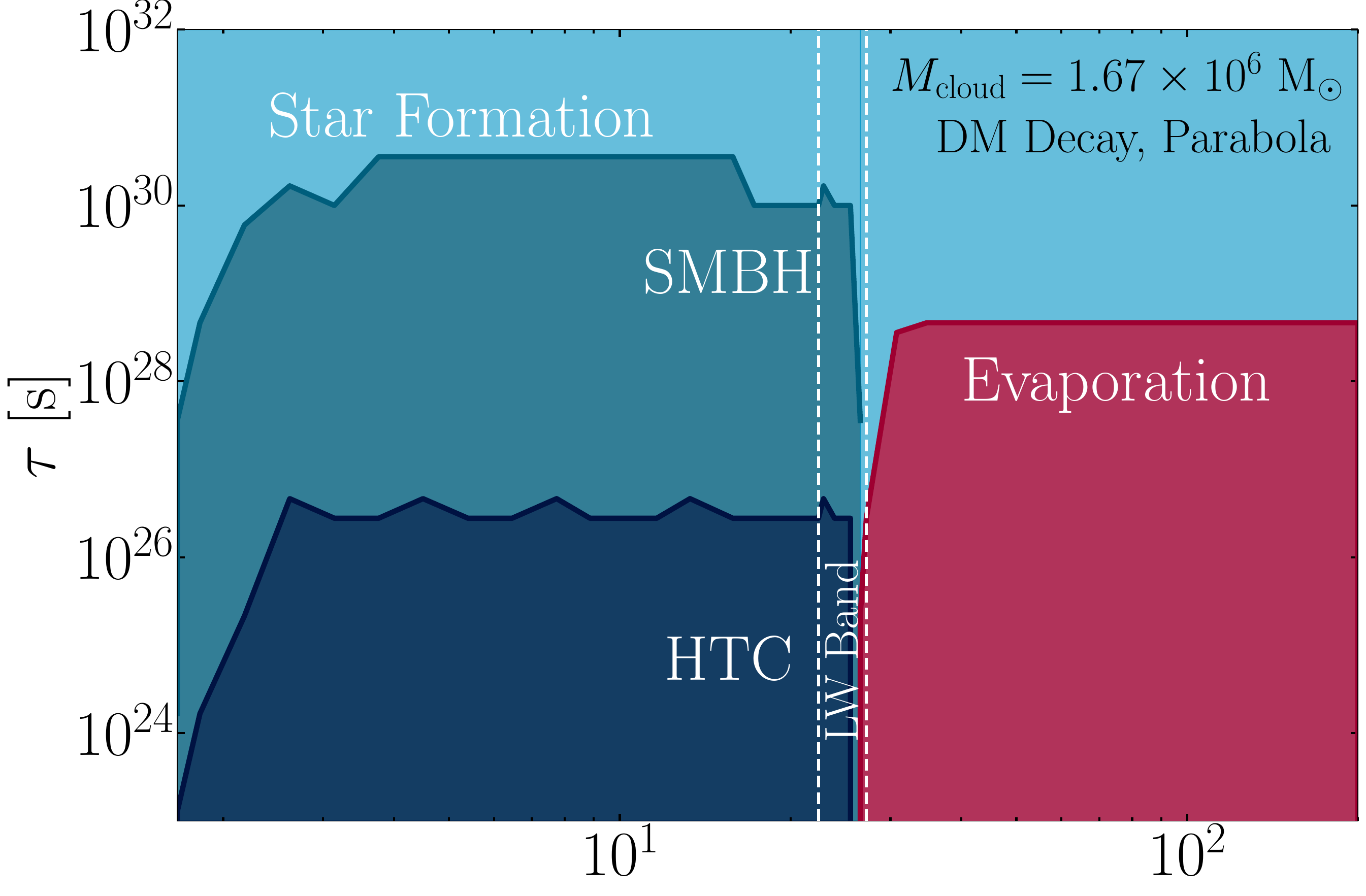} \\
\includegraphics[width=0.99\columnwidth]{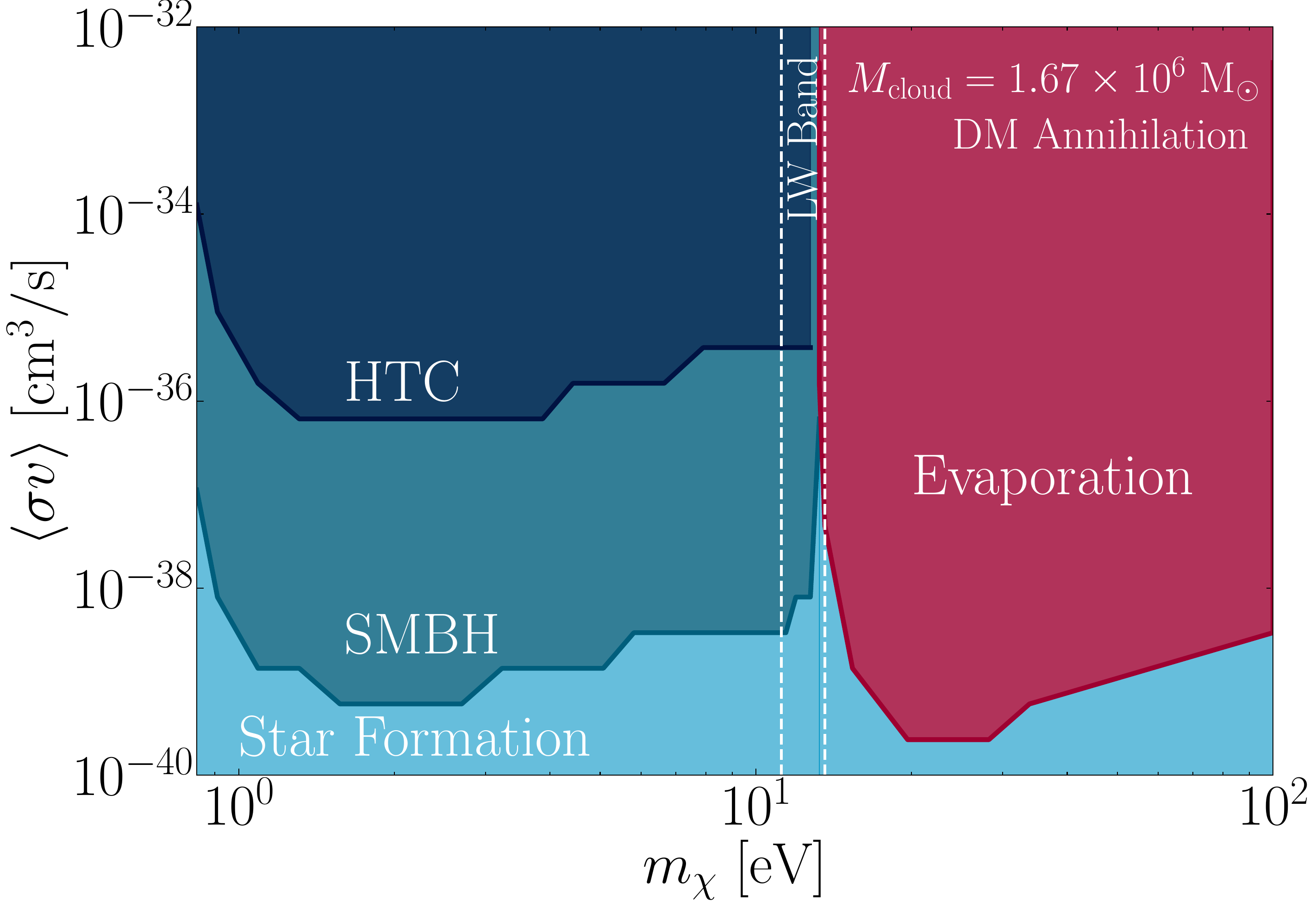} \\
\caption{Phase diagrams of the gas cloud evolution as a function of the DM parameters when self-shielding is not accounted for. White dashed lines give the mass range of DM that produces photons with energy in the LW band. The mass of the gas cloud is $1.67\times 10^6\ \rm M_\odot$. Evaporation contours appear here due to the overestimation of the high energy photon flux injection, leading to an unphysical parametrization.  {\it Top:} DM decay with a box spectrum. {\it Middle:} DM decay with a parabola-shaped photon spectrum. {\it Bottom} Non-thermal Majorana DM annihilation.\label{fig:Noss_fig}}
\end{figure}

\begin{figure}[!htb]
\centering
\includegraphics[width=0.98\columnwidth]{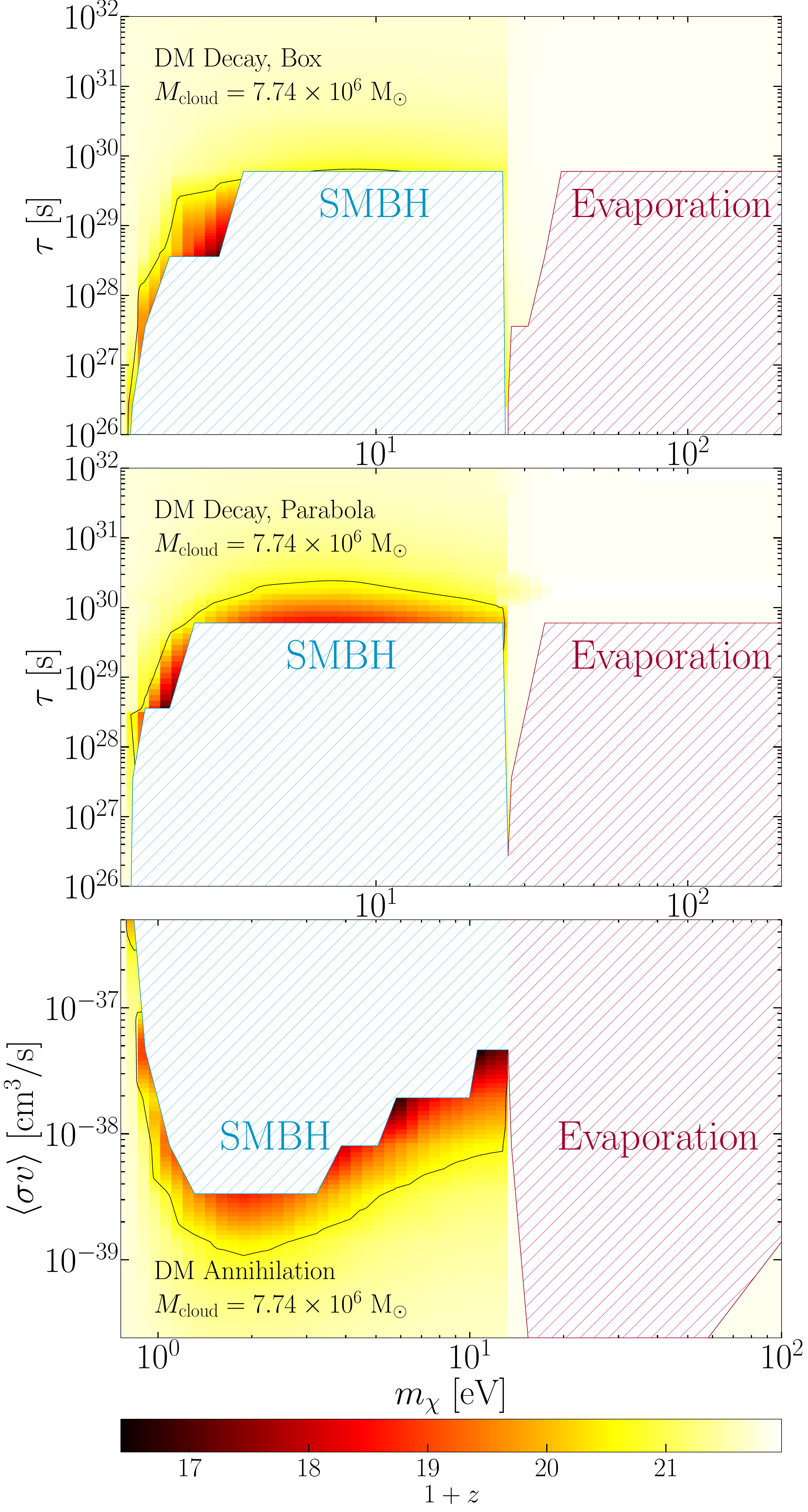}
\caption{Phase diagrams of the gas cloud evolution as a function of the DM parameters when self-shielding is not accounted for. Black curves represent the boundaries of contours which star formation not later than $5\%$ of star formation time of SM (no DM) case.  Evaporation contours appear here due to the overestimation of the high energy photon flux injection, leading to an unphysical parametrization.  {\it Top:} DM decay with a box spectrum. {\it Middle:} DM decay with a parabola-shaped photon spectrum. {\it Bottom} Non-thermal Majorana DM annihilation where. \label{fig:star_formation_noss}}
\end{figure}

In Fig.~\ref{fig:Noss_fig} and Fig.~\ref{fig:star_formation_noss}, we present the DM phase diagrams for $1.67\times10^6 \ \rm M_\odot$ gas cloud evolution without self-shielding effect and the star formation redshift for $7.74\times10^6\ \rm M_\odot$ gas cloud respectively. In fig~\ref{fig:Noss_fig}, as in the self- shielded case, increasing the injected photon flux below 13.6 eV suppresses $\rm H_2$, slowing the cooling after virialization and leading to SMBH seed or, for stronger injection, HTC formation. Unlike the self-shielded case, photons above the H ionization threshold now ionize the neutral hydrogen and raise $x_e$ at high redshift. In our simulations this is followed by a rapid temperature drop, which we label evaporation.

In Fig.~\ref{fig:star_formation_noss}, as in the self-shielded case, star formation is delayed near the boundary between star and SMBH seed formation when the photon energies lie below 13.6 eV. For higher-energy injection, however, the elevated $x_e$ accelerates cooling after virialization, and therefore star formation.\\

\section{Chemical Reaction Cross Sections}\label{chemical reactions}
In this section we provide the fitting of cross sections of chemical reactions which are dependent with external photon injection in the gas cloud from references.
\subsubsection{$\rm H,\ He,\ He^+$ Photoionization}
The fitting cross section of $\rm H,\ He$ and $\rm He^+ $ photoionization reactions is~\cite{Verner:1996th} 
\begin{equation}
    \sigma_{\rm ion}=\sigma_0 F(y)\ \rm Mb,
\end{equation}
where
\begin{equation}
    x=E/E_0-y_0,\ y=\sqrt{x^2+y_1^2},
\end{equation}
and 
\begin{equation}
    F(y)=\left[\left(x-1\right)^2+y_w^2\right]y^{0.5P-5.5}\left(1+\sqrt{y/y_a}\right)^{-P}.
\end{equation}
$E$ denotes the energy of photon, the values of $E_0$, $\sigma_0$, $y_a$, $P$, $y_w$, $y_0$ and $y_1$ are given in Table.~\ref{tab:photoionization}.
\begin{table}[!h]

\caption{Fitting parameters of $\rm H$, $\rm He$ and $\rm He^+$ photoionization reactions.\label{tab:photoionization}}
\begin{tabular}{ccccccccc}
\hline
           & $E_{\rm th}$ & $E_0\ (\rm eV)$ & $\sigma_0\ (\rm Mb)$ & $y_a$ & $P$   & $y_w$ & $y_0$ & $y_1$ \\ \hline
$\rm H$    & 13.60        & 0.4298          & $5.475\times10^4$    & 32.88 & 2.963 & 0     & 0     & 0     \\
$\rm He$   & 24.59        & 13.61          & $9.492\times10^2$    & 1.469 & 3.188 & 2.039 & 0.443 & 2.136 \\
$\rm He^+$ & 54.42        & 1.72            & $1.369\times10^4$    & 32.88 & 2.963 & 0     & 0     & 0     \\ \hline
\end{tabular}
\end{table}
\subsubsection{$\rm H_2$ Photoionization $\rm H_2+\gamma\rightarrow\rm H_2^+ +e^-$}
The cross section of $\rm H_2$ photoionization can be fitted as~\cite{yan1998photoionization}:
\begin{widetext}
\begin{equation}
\sigma_{\mathrm{H}_2} =
\begin{cases}
10^{8}\left(-37.895 + 99.723x - 87.277x^2 + 25.400x^3\right), & 15.4 \le E < 18~\text{eV},\\
2\times10^{7}\left(0.071x^{-s} - 0.673x^{-(s+1)} + 1.977x^{-(s+2)} - 0.692x^{-(s+3)}\right), & 18 \le E < 85~\text{eV},\\
45.57\left(1 - \frac{2.033}{x^{0.5}} - \frac{4.806}{x} + \frac{50.577}{x^{1.5}} - \frac{170.044}{x^2} + \frac{231.608}{x^{2.5}} - \frac{81.885}{x^3}\right), & E \ge 85~\text{eV},
\end{cases}
\end{equation}
\end{widetext}
Where $E$ is the energy of photon, $x=E(\rm eV)/15.4$, $s=0.252$, the unit of cross section is in barn.
\subsubsection{$\rm H_2^+ +\gamma\rightarrow\rm H+H^+$}
The fitting cross section of reaction is~\cite{shapiro1987hydrogen}
\begin{equation}
    \sigma_{\rm H^+_2}=\begin{cases}

10^{-40.97+6.03E-0.504E^2+1.387\times 10^{-2}E^3} & \\ \qquad\qquad\qquad\qquad2.65\le E< 11.27\ \rm eV,\\

10^{-30.26+2.79E-0.184E^2+3.535\times10^{-3}E^3} & \\
\qquad\qquad\qquad\qquad11.27\le E< 21.0\ \rm eV.

\end{cases}
\end{equation}
Where $E$ is the energy of photon, the unit of cross section is in $\rm cm^2$.
\subsubsection{$\rm H_2^+ +\gamma\rightarrow\rm 2H^+ +e^-$}
The fitting cross section of reaction is~\cite{shapiro1987hydrogen}
\begin{equation}
    \sigma_{\rm H_2}^*=10^{-16.926-4.528\times10^{-2}E+2.238\times10^{-3}E^2+4.245\times10^{-7}E^3},
\end{equation}
where $E$ is the energy of photon, the energy threshold of this reaction is $30\ \rm eV\le E \le 100\ \rm eV$, the unit of cross section is in $\rm cm^2$.

\vspace{1cm}

\bibliographystyle{apsrev4-1}
\bibliography{references}

@article{Lu:2023xoi,
    author = "Lu, Yifan and Picker, Zachary S. C. and Kusenko, Alexander",
    title = "{High-redshift supermassive black holes from tiny black hole explosions}",
    eprint = "2312.15062",
    archivePrefix = "arXiv",
    primaryClass = "astro-ph.GA",
    reportNumber = "IPMU23-0052",
    doi = "10.1103/PhysRevD.109.123016",
    journal = "Phys. Rev. D",
    volume = "109",
    number = "12",
    pages = "123016",
    year = "2024"
}

@article{shapiro1987hydrogen,
  title={Hydrogen molecules and the radiative cooling of pregalactic shocks},
  author={Shapiro, Paul R and Kang, Hyesung},
  journal={The Astrophysical Journal},
  url ={https://ui.adsabs.harvard.edu/abs/1987ApJ...318...32S/abstract},
  volume={318},
  pages={32--65},
  year={1987}
}

@article{wolcott2017beyond,
    author = "Wolcott-Green, J. and Haiman, Z. and Bryan, G. L.",
    title = "{Beyond Jcrit: a critical curve for suppression of H2-cooling in protogalaxies}",
    eprint = "1609.02142",
    archivePrefix = "arXiv",
    primaryClass = "astro-ph.GA",
    doi = "10.1093/mnras/stx167",
    journal = "Mon. Not. Roy. Astron. Soc.",
    volume = "469",
    number = "3",
    pages = "3329--3336",
    year = "2017"
}

@article{Omukai:2000ic,
    author = "Omukai, Kazuyuki",
    title = "{Primordial star formation under far-ultraviolet radiation}",
    eprint = "astro-ph/0011446",
    archivePrefix = "arXiv",
    reportNumber = "NAOJ-TH-AP-2000",
    doi = "10.1086/318296",
    journal = "Astrophys. J.",
    volume = "546",
    pages = "635",
    year = "2001"
}

@ARTICLE{Schon2018,
       author = {{Sch{\"o}n}, S. and {Mack}, K.~J. and {Wyithe}, J.~S.~B.},
        title = "{Dark matter annihilation in the circumgalactic medium at high redshifts}",
      journal = {\mnras},
         year = 2018,
        month = mar,
       volume = {474},
       number = {3},
        pages = {3067-3079},
          doi = {10.1093/mnras/stx2968},
archivePrefix = {arXiv},
       eprint = {1706.04327},
 primaryClass = {astro-ph.CO},
       adsurl = {https://ui.adsabs.harvard.edu/abs/2018MNRAS.474.3067S}
}

@article{Liu:2023nct,
    author = "Liu, Hongwan and Qin, Wenzer and Ridgway, Gregory W. and Slatyer, Tracy R.",
    title = "{Exotic energy injection in the early Universe. II. CMB spectral distortions and constraints on light dark matter}",
    eprint = "2303.07370",
    archivePrefix = "arXiv",
    primaryClass = "astro-ph.CO",
    reportNumber = "MIT-CTP/5524",
    doi = "10.1103/PhysRevD.108.043531",
    journal = "Phys. Rev. D",
    volume = "108",
    number = "4",
    pages = "043531",
    year = "2023"
}

@article{Liu:2023fgu,
    author = "Liu, Hongwan and Qin, Wenzer and Ridgway, Gregory W. and Slatyer, Tracy R.",
    title = "{Exotic energy injection in the early Universe. I. A novel treatment for low-energy electrons and photons}",
    eprint = "2303.07366",
    archivePrefix = "arXiv",
    primaryClass = "astro-ph.CO",
    reportNumber = "MIT-CTP/5523",
    doi = "10.1103/PhysRevD.108.043530",
    journal = "Phys. Rev. D",
    volume = "108",
    number = "4",
    pages = "043530",
    year = "2023"
}

@article{Ripamonti:2006gr,
    author = "Ripamonti, E. and Mapelli, Michela and Ferrara, A.",
    title = "{The impact of dark matter decays and annihilations on the formation of the first structures}",
    eprint = "astro-ph/0606483",
    archivePrefix = "arXiv",
    doi = "10.1111/j.1365-2966.2006.11402.x",
    journal = "Mon. Not. Roy. Astron. Soc.",
    volume = "375",
    pages = "1399--1408",
    year = "2007"
}

@article{Cyr:2022urs,
    author = "Cyr, Bryce and Jiao, Hao and Brandenberger, Robert",
    title = "{Massive black holes at high redshifts from superconducting cosmic strings}",
    eprint = "2202.01799",
    archivePrefix = "arXiv",
    primaryClass = "astro-ph.CO",
    doi = "10.1093/mnras/stac1939",
    journal = "Mon. Not. Roy. Astron. Soc.",
    volume = "517",
    number = "2",
    pages = "2221--2230",
    year = "2022"
}

@article{Hirano:2017wbu,
    author = "Hirano, Shingo and Hosokawa, Takashi and Yoshida, Naoki and Kuiper, Rolf",
    title = "{Supersonic Gas Streams Enhance the Formation of Massive Black Holes in the Early Universe}",
    eprint = "1709.09863",
    archivePrefix = "arXiv",
    primaryClass = "astro-ph.CO",
    doi = "10.1126/science.aai9119",
    journal = "Science",
    volume = "357",
    pages = "1375",
    year = "2017"
}

@article{navarro1997universal,
  title={A universal density profile from hierarchical clustering},
  author={Navarro, Julio F and Frenk, Carlos S and White, Simon DM},
  eprint = "arXiv:astro-ph/9611107",
  archivePrefix = "arXiv",
  journal={The Astrophysical Journal},
  volume={490},
  number={2},
  pages={493},
  year={1997},
  publisher={IOP Publishing}
}

@article{bullock2001profiles,
  title={Profiles of dark haloes: evolution, scatter and environment},
  author={Bullock, James S and Kolatt, Tsafrir S and Sigad, Yair and Somerville, Rachel S and Kravtsov, Andrey V and Klypin, Anatoly A and Primack, Joel R and Dekel, Avishai},
  eprint = "arXiv:astro-ph/9908159",
  archivePrefix = "arXiv",
  journal={Monthly Notices of the Royal Astronomical Society},
  volume={321},
  number={3},
  pages={559--575},
  year={2001},
  publisher={Blackwell Science Ltd Oxford, UK}
}

@article{jing2000density,
  title={The Density Profile of Equilibrium and Nonequilibrium DarkMatter Halos},
  author={Jing, YP},
  eprint = "astro-ph/9901340",
  archivePrefix = "arXiv",
  journal={The Astrophysical Journal},
  volume={535},
  number={1},
  pages={30},
  year={2000},
  publisher={IOP Publishing}
}

@article{Wolcott-Green:2011tul,
    author = "Wolcott-Green, Jemma and Haiman, Zoltan and Bryan, Greg L.",
    title = "{Photodissociation of H2 in Protogalaxies: Modeling Self-Shielding in 3D Simulations}",
    eprint = "1106.3523",
    archivePrefix = "arXiv",
    primaryClass = "astro-ph.CO",
    doi = "10.1111/j.1365-2966.2011.19538.x",
    journal = "Mon. Not. Roy. Astron. Soc.",
    volume = "418",
    pages = "838",
    year = "2011"
}

@article{Chardin:2017juq,
    author = "Chardin, Jonathan and Kulkarni, Girish and Haehnelt, Martin G.",
    title = "{Self-shielding of hydrogen in the IGM during the epoch of reionization}",
    eprint = "1707.06993",
    archivePrefix = "arXiv",
    primaryClass = "astro-ph.CO",
    doi = "10.1093/mnras/sty992",
    journal = "Mon. Not. Roy. Astron. Soc.",
    volume = "478",
    number = "1",
    pages = "1065--1076",
    year = "2018"
}

@article{Rahmati:2012rg,
    author = "Rahmati, Alireza and Pawlik, Andreas P. and Raicevic, Milan and Schaye, Joop",
    title = "{On the evolution of the HI column density distribution in cosmological simulations}",
    eprint = "1210.7808",
    archivePrefix = "arXiv",
    primaryClass = "astro-ph.CO",
    doi = "10.1093/mnras/stt066",
    journal = "Mon. Not. Roy. Astron. Soc.",
    volume = "430",
    pages = "2427",
    year = "2013"
}

@article{Ibarra:2012dw,
    author = "Ibarra, Alejandro and Lopez Gehler, Sergio and Pato, Miguel",
    title = "{Dark matter constraints from box-shaped gamma-ray features}",
    eprint = "1205.0007",
    archivePrefix = "arXiv",
    primaryClass = "hep-ph",
    doi = "10.1088/1475-7516/2012/07/043",
    journal = "JCAP",
    volume = "07",
    pages = "043",
    year = "2012"
}

@article{Miralda-Escude:2003lmt,
    author = "Miralda-Escude, Jordi",
    title = "{The dark age of the universe}",
    eprint = "astro-ph/0307396",
    archivePrefix = "arXiv",
    doi = "10.1126/science.1085325",
    journal = "Science",
    volume = "300",
    pages = "1904--1909",
    year = "2003"
}

@article{Rees:1999kna,
    author = "Rees, Martin J.",
    title = "{`first light' in the universe: what ended the dark age?}",
    eprint = "astro-ph/9912345",
    archivePrefix = "arXiv",
    doi = "10.1016/S0370-1573(00)00023-5",
    journal = "Phys. Rept.",
    volume = "333",
    pages = "203",
    year = "2000"
}

@article{Rees:1996ij,
    author = "Rees, Martin J.",
    title = "{The Universe at Z {\ensuremath{>}} 5: When and how did the 'dark age' end?}",
    eprint = "astro-ph/9608196",
    archivePrefix = "arXiv",
    month = "8",
    year = "1996"
}

@article{Jester:2009dw,
    author = "Jester, Sebastian and Falcke, Heino",
    title = "{Science with a lunar low-frequency array: from the dark ages of the Universe to nearby exoplanets}",
    eprint = "0902.0493",
    archivePrefix = "arXiv",
    primaryClass = "astro-ph.CO",
    doi = "10.1016/j.newar.2009.02.001",
    journal = "New Astron. Rev.",
    volume = "53",
    pages = "1--26",
    year = "2009"
}

@article{Parijskij:1998ycu,
    author = "Parijskij, Y. and Kopylov, A. and Soboleva, N. and Verkhodanov, O. and Temirova, A. and Zhelenkova, O. and Tcibulev, P. and Chepurnov, A. and Stolyarov, V. and Bursov, N.",
    editor = "S{\'a}nchez, N. and Zichichi, A.",
    title = "{{\textquotedblright}Dark Ages{\textquotedblright} of the Universe}",
    doi = "10.1007/978-94-011-5046-0_14",
    journal = "NATO Sci. Ser. C",
    volume = "511",
    pages = "443--466",
    year = "1998"
}

@article{Bertschinger:1998tv,
    author = "Bertschinger, Edmund",
    title = "{Simulations of structure formation in the universe}",
    doi = "10.1146/annurev.astro.36.1.599",
    journal = "Ann. Rev. Astron. Astrophys.",
    volume = "36",
    pages = "599--654",
    year = "1998"
}

@article{primack1997dark,
  title={Dark matter and structure formation in the Universe},
  author={Primack, Joel R},
  journal={arXiv preprint astro-ph/9707285},
  eprint="astro-ph/9707285",
  archivePrefix = "arXiv",
  year={1997}
}

@article{Friedlander:2022ovf,
    author = "Friedlander, Avi and Schon, Sarah and Vincent, Aaron C.",
    title = "{Supermassive black hole seeds from sub-keV dark matter}",
    eprint = "2212.11100",
    archivePrefix = "arXiv",
    primaryClass = "hep-ph",
    doi = "10.1088/1475-7516/2023/06/033",
    journal = "JCAP",
    volume = "06",
    pages = "033",
    year = "2023"
}

@article{Gnedin:2000uj,
    author = "Gnedin, Nickolay Y.",
    title = "{Effect of reionization on the structure formation in the universe}",
    eprint = "astro-ph/0002151",
    archivePrefix = "arXiv",
    doi = "10.1086/317042",
    journal = "Astrophys. J.",
    volume = "542",
    pages = "535--541",
    year = "2000"
}

@book{loeb2013first,
  title={The first galaxies in the universe},
  author={Loeb, Abraham and Furlanetto, Steven R},
  year={2013},
  publisher={Princeton University Press},
  doi = {10.23943/princeton/9780691144917.001.0001},
  url = {https://doi.org/10.23943/princeton/9780691144917.001.0001}
}

@article{Dietrich:2001fc,
    author = "Dietrich, M. and Appenzeller, I. and Vestergaard, M. and Wagner, S. J.",
    title = "{High redshift quasars and star formation in the early universe}",
    eprint = "astro-ph/0109208",
    archivePrefix = "arXiv",
    doi = "10.1086/324337",
    journal = "Astrophys. J.",
    volume = "564",
    pages = "581",
    year = "2002"
}

@article{Robertson:2022gdk,
    author = "Robertson, B. E. and others",
    title = "{Identification and properties of intense star-forming galaxies at redshifts z{\,}{\ensuremath{>}}{\,}10}",
    eprint = "2212.04480",
    archivePrefix = "arXiv",
    primaryClass = "astro-ph.GA",
    doi = "10.1038/s41550-023-01921-1",
    url = "https://arxiv.org/abs/2212.04480",
    journal = "Nature Astron.",
    volume = "7",
    number = "5",
    pages = "611--621",
    year = "2023"
}

@article{bradley2023high,
  title={High-redshift Galaxy Candidates at z= 9--10 as Revealed by JWST Observations of WHL0137-08},
  author={Bradley, Larry D and Coe, Dan and Brammer, Gabriel and Furtak, Lukas J and Larson, Rebecca L and Kokorev, Vasily and Andrade-Santos, Felipe and Bhatawdekar, Rachana and Brada{\v{c}}, Maru{\v{s}}a and Broadhurst, Tom and others},
  eprint = "2210.01777",
  archivePrefix = "arXiv",
  journal={The Astrophysical Journal},
  volume={955},
  number={1},
  pages={13},
  year={2023},
  publisher={The American Astronomical Society}
}

@article{adams2023discovery,
  title={Discovery and properties of ultra-high redshift galaxies (9< z< 12) in the JWST ERO SMACS 0723 Field},
  author={Adams, Nathan J and Conselice, Christopher J and Ferreira, Leonardo and Austin, Duncan and Trussler, James AA and Juod{\v{z}}balis, Ignas and Wilkins, SM and Caruana, J and Dayal, P and Verma, A and others},
  eprint = "2207.11217 ",
  archivePrefix = "arXiv",
  primaryClass = "astro-ph.GA",
  journal={Monthly Notices of the Royal Astronomical Society},
  volume={518},
  number={3},
  pages={4755--4766},
  year={2023},
  publisher={Oxford University Press}
}

@article{Ciardi:2015oia,
    author = "Ciardi, Benedetta and Inoue, Susumu and Mack, Katherine J. and Xu, Yidong and Bernardi, Gianni",
    editor = "Bourke, Tyler L. and others",
    title = "{21cm Forest with the SKA}",
    eprint = "1501.04425",
    archivePrefix = "arXiv",
    primaryClass = "astro-ph.CO",
    doi = "10.22323/1.215.0006",
    journal = "PoS",
    volume = "AASKA14",
    pages = "006",
    year = "2015"
}

@article{HERA:2025ajm,
    author = "Abdurashidova, Zuhra and others",
    collaboration = "HERA",
    title = "{First Results from HERA Phase II}",
    eprint = "2511.21289",
    archivePrefix = "arXiv",
    primaryClass = "astro-ph.CO",
    doi = "10.3847/1538-4357/ae2d54",
    journal = "Astrophys. J.",
    volume = "998",
    number = "1",
    pages = "33",
    year = "2026"
}

@article{Garsden:2021kdo,
    author = "Garsden, Hugh and Greenhill, Lincoln and Bernardi, Gianni and Fialkov, Anastasia and Price, Daniel C. and Mitchell, Daniel and Dowell, Jayce and Spinelli, Marta and Schinzel, Frank K.",
    title = "{A 21-cm power spectrum at 48~MHz, using the Owens Valley Long Wavelength Array}",
    eprint = "2102.09596",
    archivePrefix = "arXiv",
    primaryClass = "astro-ph.CO",
    doi = "10.1093/mnras/stab1671",
    journal = "Mon. Not. Roy. Astron. Soc.",
    volume = "506",
    number = "4",
    pages = "5802--5817",
    year = "2021"
}

@article{Voytek:2013nua,
    author = "Voytek, Tabitha C. and Natarajan, Aravind and J{\'a}uregui Garc{\'\i}a, Jos{\'e} Miguel and Peterson, Jeffrey B. and L{\'o}pez-Cruz, Omar",
    title = "{Probing the Dark Ages at $z \sim$ 20: The SCI-HI 21 cm All-sky Spectrum Experiment}",
    eprint = "1311.0014",
    archivePrefix = "arXiv",
    primaryClass = "astro-ph.CO",
    doi = "10.1088/2041-8205/782/1/L9",
    journal = "Astrophys. J. Lett.",
    volume = "782",
    pages = "L9",
    year = "2014"
}

@article{Goulding:2023gqa,
    author = "Goulding, Andy D. and others",
    title = "{UNCOVER: The Growth of the First Massive Black Holes from JWST/NIRSpec{\textemdash}Spectroscopic Redshift Confirmation of an X-Ray Luminous AGN at z = 10.1}",
    eprint = "2308.02750",
    archivePrefix = "arXiv",
    primaryClass = "astro-ph.GA",
    doi = "10.3847/2041-8213/acf7c5",
    journal = "Astrophys. J. Lett.",
    volume = "955",
    number = "1",
    pages = "L24",
    year = "2023"
}

@article{Menci:2022wia,
    author = "Menci, N. and Castellano, M. and Santini, P. and Merlin, E. and Fontana, A. and Shankar, F.",
    title = "{High-redshift Galaxies from Early JWST Observations: Constraints on Dark Energy Models}",
    eprint = "2208.11471",
    archivePrefix = "arXiv",
    primaryClass = "astro-ph.CO",
    doi = "10.3847/2041-8213/ac96e9",
    journal = "Astrophys. J. Lett.",
    volume = "938",
    number = "1",
    pages = "L5",
    year = "2022"
}

@article{Matthee:2023utn,
    author = "Matthee, Jorryt and others",
    title = "{Little Red Dots: An Abundant Population of Faint Active Galactic Nuclei at z {\ensuremath{\sim}} 5 Revealed by the EIGER and FRESCO JWST Surveys}",
    eprint = "2306.05448",
    archivePrefix = "arXiv",
    primaryClass = "astro-ph.GA",
    doi = "10.3847/1538-4357/ad2345",
    journal = "Astrophys. J.",
    volume = "963",
    number = "2",
    pages = "129",
    year = "2024"
}

@article{Haiman:2004ve,
    author = "Haiman, Zoltan",
    title = "{Constraints from gravitational recoil on the growth of supermassive black holes at high redshift}",
    eprint = "astro-ph/0404196",
    archivePrefix = "arXiv",
    doi = "10.1086/422910",
    journal = "Astrophys. J.",
    volume = "613",
    pages = "36--40",
    year = "2004"
}

@article{Tanaka:2008bv,
    author = "Tanaka, Takamitsu and Haiman, Zoltan",
    title = "{The Assembly of Supermassive Black Holes at High Redshifts}",
    eprint = "0807.4702",
    archivePrefix = "arXiv",
    primaryClass = "astro-ph",
    doi = "10.1088/0004-637X/696/2/1798",
    journal = "Astrophys. J.",
    volume = "696",
    pages = "1798--1822",
    year = "2009"
}

@article{Qin:2023kkk,
    author = "Qin, Wenzer and Munoz, Julian B. and Liu, Hongwan and Slatyer, Tracy R.",
    title = "{Birth of the first stars amidst decaying and annihilating dark matter}",
    eprint = "2308.12992",
    archivePrefix = "arXiv",
    primaryClass = "astro-ph.CO",
    reportNumber = "MIT-CTP/5596",
    doi = "10.1103/PhysRevD.109.103026",
    journal = "Phys. Rev. D",
    volume = "109",
    number = "10",
    pages = "103026",
    year = "2024"
}

@article{Lu:2024zwa,
    author = "Lu, Yifan and Picker, Zachary S. C. and Kusenko, Alexander",
    title = "{Direct Collapse Supermassive Black Holes from Relic Particle Decay}",
    eprint = "2404.03909",
    archivePrefix = "arXiv",
    primaryClass = "astro-ph.GA",
    reportNumber = "IPMU24-0013",
    doi = "10.1103/PhysRevLett.133.091001",
    journal = "Phys. Rev. Lett.",
    volume = "133",
    number = "9",
    pages = "091001",
    year = "2024"
}

@article{Tegmark:1996yt,
    author = "Tegmark, Max and Silk, Joseph and Rees, Martin J. and Blanchard, Alain and Abel, Tom and Palla, Francesco",
    title = "{How small were the first cosmological objects?}",
    eprint = "astro-ph/9603007",
    archivePrefix = "arXiv",
    doi = "10.1086/303434",
    journal = "Astrophys. J.",
    volume = "474",
    pages = "1--12",
    year = "1997"
}

@article{Zhao:2002rm,
    author = "Zhao, Donghai and Mo, Houjun and Jing, Yipeng and Boerner, Gerhard",
    title = "{The growth and structure of dark matter haloes}",
    eprint = "arXiv:astro-ph/0204108",
    archivePrefix = "arXiv",
    doi = "10.1046/j.1365-8711.2003.06135.x",
    journal = "Mon. Not. Roy. Astron. Soc.",
    volume = "339",
    pages = "12--24",
    year = "2003"
}

@article{zhao2003mass,
  title={Mass and redshift dependence of dark halo structure},
  author={Zhao, Dong-Hai and Jing, YP and Mo, HJ and B{\"o}rner, G},
  eprint = "arXiv:astro-ph/0309375",
  archivePrefix = "arXiv",
  journal={The Astrophysical Journal Letters},
  volume={597},
  number={1},
  pages={L9--L12},
  year={2003}
}

@article{Li:2005jm,
    author = "Li, Yun and Mo, H. J. and van den Bosch, Frank C. and Lin, W. P.",
    title = "{The Assembly History of Dark Matter Haloes}",
    eprint = "astro-ph/0510372",
    archivePrefix = "arXiv",
    doi = "10.1111/j.1365-2966.2007.11942.x",
    journal = "Mon. Not. Roy. Astron. Soc.",
    volume = "379",
    pages = "689--701",
    year = "2007"
}

@article{Hoffman:2007tu,
    author = "Hoffman, Yehuda and Romano-Diaz, Emilio and Shlosman, Isaac and Heller, Clayton",
    title = "{Evolution of Phase-Space Density in Dark Matter Halos}",
    eprint = "0706.0006",
    archivePrefix = "arXiv",
    primaryClass = "astro-ph",
    doi = "10.1086/523695",
    journal = "Astrophys. J.",
    volume = "671",
    pages = "1108",
    year = "2007"
}

@article{jeans1902stability,
  title={I. The stability of a spherical nebula},
  author={Jeans, James Hopwood},
  journal={Philosophical Transactions of the Royal Society of London. Series A, Containing Papers of a Mathematical or Physical Character},
  volume={199},
  number={312-320},
  pages={1--53},
  year={1902},
  publisher={The Royal Society London},
  doi = {10.1098/rsta.1902.0012},
  url = {https://doi.org/10.1098/rsta.1902.0012}
}

@article{Barkana:2000fd,
    author = "Barkana, Rennan and Loeb, Abraham",
    title = "{In the beginning: The First sources of light and the reionization of the Universe}",
    eprint = "astro-ph/0010468",
    archivePrefix = "arXiv",
    doi = "10.1016/S0370-1573(01)00019-9",
    journal = "Phys. Rept.",
    volume = "349",
    pages = "125--238",
    year = "2001"
}

@article{Smith:2016hsc,
    author = "Smith, Britton D. and others",
    title = "{Grackle: a Chemistry and Cooling Library for Astrophysics}",
    eprint = "1610.09591",
    archivePrefix = "arXiv",
    primaryClass = "astro-ph.CO",
    doi = "10.1093/mnras/stw3291",
    journal = "Mon. Not. Roy. Astron. Soc.",
    volume = "466",
    number = "2",
    pages = "2217--2234",
    year = "2017"
}

@article{Lesgourgues:2011rg,
    author = "Lesgourgues, Julien",
    title = "{The Cosmic Linear Anisotropy Solving System (CLASS) III: Comparision with CAMB for LambdaCDM}",
    eprint = "1104.2934",
    archivePrefix = "arXiv",
    primaryClass = "astro-ph.CO",
    reportNumber = "CERN-PH-TH-2011-083, LAPTH-011-11",
    month = "4",
    year = "2011"
}

@article{Haardt:2011xv,
    author = "Haardt, Francesco and Madau, Piero",
    title = "{Radiative transfer in a clumpy universe: IV. New synthesis models of the cosmic UV/X-ray background}",
    eprint = "1105.2039",
    archivePrefix = "arXiv",
    primaryClass = "astro-ph.CO",
    doi = "10.1088/0004-637X/746/2/125",
    journal = "Astrophys. J.",
    volume = "746",
    pages = "125",
    year = "2012"
}

@article{Verner:1996th,
    author = "Verner, D. A. and Ferland, G. J. and Korista, K. T. and Yakovlev, D. G.",
    title = "{Atomic data for astrophysics. 2. New analytic FITS for photoionization cross-sections of atoms and ions}",
    eprint = "astro-ph/9601009",
    archivePrefix = "arXiv",
    doi = "10.1086/177435",
    journal = "Astrophys. J.",
    volume = "465",
    pages = "487",
    year = "1996"
}

@article{yan1998photoionization,
  title={Photoionization cross sections of He and H2},
  author={Yan, M and Sadeghpour, HR and Dalgarno, A},
  journal={The Astrophysical Journal},
  volume={496},
  number={2},
  pages={1044--1050},
  year={1998},
  url={https://ui.adsabs.harvard.edu/abs/1998ApJ...496.1044Y/abstract}
}

@article{Bhattacharya:2025dgx,
    author = "Bhattacharya, Sulagna and Bose, Debajit and Dasgupta, Basudeb and Doliya, Jaya and Laha, Ranjan",
    title = "{Dark Recipe for the First Giants: From Population III Stars to Early Supermassive Black Holes via Dark Matter Capture}",
    eprint = "2512.23789",
    archivePrefix = "arXiv",
    primaryClass = "hep-ph",
    month = "12",
    year = "2025"
}

@article{Willott:2010yu,
    author = "Willott, Chris J. and Albert, Loic and Arzoumanian, Doris and Bergeron, Jacqueline and Crampton, David and Delorme, Philippe and Hutchings, John B. and Omont, Alain and Reyle, Celine and Schade, David",
    title = "{Eddington-limited accretion and the black hole mass function at redshift 6}",
    eprint = "1006.1342",
    archivePrefix = "arXiv",
    primaryClass = "astro-ph.CO",
    doi = "10.1088/0004-6256/140/2/546",
    journal = "Astron. J.",
    volume = "140",
    pages = "546",
    year = "2010"
}

@article{Wu:2015phw,
    author = "Wu, Xue-Bing and others",
    title = "{An ultraluminous quasar with a twelve-billion-solar-mass black hole at redshift 6.30}",
    doi = "10.1038/nature14241",
    journal = "Nature",
    volume = "518",
    number = "7540",
    pages = "512--515",
    year = "2015"
}

@article{Banados:2017unc,
    author = "Banados, Eduardo and others",
    title = "{An 800-million-solar-mass black hole in a significantly neutral Universe at redshift 7.5}",
    eprint = "1712.01860",
    archivePrefix = "arXiv",
    primaryClass = "astro-ph.GA",
    doi = "10.1038/nature25180",
    journal = "Nature",
    volume = "553",
    number = "7689",
    pages = "473--476",
    year = "2018"
}

@article{Oesch:2016dys,
    author = "Oesch, P. A. and others",
    title = "{A Remarkably Luminous Galaxy at Z~=~11.1 Measured With Hubble Space Telescope Grism Spectroscopy}",
    doi = "10.3847/0004-637x/819/2/129",
    journal = "Astrophys. J.",
    volume = "819",
    number = "2",
    pages = "129",
    year = "2016"
}

@article{napolitano2025dual,
  title={The Dual Nature of GHZ9: Coexisting Active Galactic Nuclei and Star Formation Activity in a Remote X-Ray Source at z= 10.145},
  author={Napolitano, Lorenzo and Castellano, Marco and Pentericci, Laura and Vignali, Cristian and Gilli, Roberto and Fontana, Adriano and Santini, Paola and Treu, Tommaso and Calabr{\`o}, Antonello and Llerena, Mario and others},
  eprint = "2410.18763",
  archivePrefix = "arXiv",
  primaryClass = "astro-ph.GA",
  journal={The Astrophysical Journal},
  volume={989},
  number={1},
  pages={75},
  year={2025},
  publisher={The American Astronomical Society}
}

@article{Taylor:2025ubx,
    author = "Taylor, Anthony J. and others",
    title = "{CAPERS-LRD-z9: A Gas-enshrouded Little Red Dot Hosting a Broad-line Active Galactic Nucleus at z = 9.288}",
    doi = "10.3847/2041-8213/ade789",
    journal = "Astrophys. J. Lett.",
    volume = "989",
    number = "1",
    pages = "L7",
    year = "2025"
}

@article{larson2023ceers,
  title={A CEERS discovery of an accreting supermassive black hole 570 Myr after the Big Bang: identifying a progenitor of massive z> 6 quasars},
  author={Larson, Rebecca L and Finkelstein, Steven L and Kocevski, Dale D and Hutchison, Taylor A and Trump, Jonathan R and Arrabal Haro, Pablo and Bromm, Volker and Cleri, Nikko J and Dickinson, Mark and Fujimoto, Seiji and others},
  eprint = "2303.08918",
  archivePrefix = "arXiv",
  primaryClass = "astro-ph.GA",
  journal={The Astrophysical Journal Letters},
  volume={953},
  number={2},
  pages={L29},
  year={2023},
  publisher={The American Astronomical Society}
}

@article{goulding2023uncover,
  title={UNCOVER: the growth of the first massive black holes from JWST/NIRSpec—spectroscopic redshift confirmation of an X-ray luminous AGN at z= 10.1},
  author={Goulding, Andy D and Greene, Jenny E and Setton, David J and Labbe, Ivo and Bezanson, Rachel and Miller, Tim B and Atek, Hakim and Bogd{\'a}n, {\'A}kos and Brammer, Gabriel and Chemerynska, Iryna and others},
  eprint = "2308.02750 ",
  archivePrefix = "arXiv",
  primaryClass = "astro-ph.GA",
  journal={The Astrophysical Journal Letters},
  volume={955},
  number={1},
  pages={L24},
  year={2023},
  publisher={The American Astronomical Society}
}

@article{furtak2023jwst,
  title={JWST UNCOVER: extremely red and compact object at z phot≃ 7.6 triply imaged by A2744},
  author={Furtak, Lukas J and Zitrin, Adi and Plat, Ad{\`e}le and Fujimoto, Seiji and Wang, Bingjie and Nelson, Erica J and Labb{\'e}, Ivo and Bezanson, Rachel and Brammer, Gabriel B and van Dokkum, Pieter and others},
  eprint = "2212.10531 ",
  archivePrefix = "arXiv",
  primaryClass = "astro-ph.GA",
  journal={The Astrophysical Journal},
  volume={952},
  number={2},
  pages={142},
  year={2023},
  publisher={The American Astronomical Society}
}

@article{lopez2013constraints,
  title={Constraints on dark matter annihilation from CMB observations before Planck},
  author={Lopez-Honorez, Laura and Mena, Olga and Palomares-Ruiz, Sergio and Vincent, Aaron C},
  eprint = "1303.5094",
  archivePrefix = "arXiv",
  primaryClass = "astro-ph.CO",
  journal={Journal of Cosmology and Astroparticle Physics},
  volume={2013},
  number={07},
  pages={046--046},
  year={2013}
}

@article{barkana2001beginning,
  title={In the beginning: the first sources of light and the reionization of the universe},
  author={Barkana, Rennan and Loeb, Abraham},
  eprint = "arXiv:astro-ph/0010468",
  archivePrefix = "arXiv",
  journal={Physics reports},
  volume={349},
  number={2},
  pages={125--238},
  year={2001},
  publisher={Elsevier}
}

@article{Hou:2024hcj,
    author = "Hou, Liqiang and Mack, Katherine J.",
    title = "{The effects of dark matter annihilation and dark matter-baryon velocity offsets at Cosmic Dawn}",
    eprint = "2411.10626",
    archivePrefix = "arXiv",
    primaryClass = "astro-ph.CO",
    doi = "10.1088/1475-7516/2025/04/081",
    journal = "JCAP",
    volume = "04",
    pages = "081",
    year = "2025"
}

@book{baumann2022cosmology,
  title={Cosmology},
  author={Baumann, Daniel},
  volume={7},
  year={2022},
  publisher={Cambridge University Press Cambridge},
  doi={10.1017/9781108937092},
  url={https://ui.adsabs.harvard.edu/abs/2022cosm.book.....B/abstract}
}

@article{Zaroubi:2012in,
    author = "Zaroubi, Saleem",
    title = "{The Epoch of Reionization}",
    eprint = "1206.0267",
    archivePrefix = "arXiv",
    primaryClass = "astro-ph.CO",
    doi = "10.1007/978-3-642-32362-1\_2",
    month = "6",
    year = "2012"
}

@article{Wise:2019qtq,
    author = "Wise, John H.",
    title = "{Cosmic reionisation}",
    eprint = "1907.06653",
    archivePrefix = "arXiv",
    primaryClass = "astro-ph.CO",
    doi = "10.1080/00107514.2019.1631548",
    journal = "Contemp. Phys.",
    volume = "60",
    number = "2",
    pages = "145--163",
    year = "2019"
}

@article{chiaki2023direct,
  title={Direct-collapse black hole formation induced by internal radiation of host haloes},
  author={Chiaki, Gen and Chon, Sunmyon and Omukai, Kazuyuki and Trinca, Alessandro and Schneider, Raffaella and Valiante, Rosa},
  eprint = "2303.01762",
  archivePrefix = "arXiv",
  primaryClass = "astro-ph.GA",
  journal={Monthly Notices of the Royal Astronomical Society},
  volume={521},
  number={2},
  pages={2845--2859},
  year={2023},
  publisher={Oxford University Press}
}

@article{dunn2018sowing,
    author = "Dunn, Glenna and Bellovary, Jillian and Holley-Bockelmann, Kelly and Christensen, Charlotte and Quinn, Thomas",
    title = "{Sowing Black Hole Seeds: Direct Collapse Black Hole Formation with Realistic Lyman{\textendash}Werner Radiation in Cosmological Simulations}",
    eprint = "1803.01007",
    archivePrefix = "arXiv",
    primaryClass = "astro-ph.GA",
    doi = "10.3847/1538-4357/aac7c2",
    journal = "Astrophys. J.",
    volume = "861",
    number = "1",
    pages = "39",
    year = "2018"
}

@article{zhang2025novel,
  title={A novel formation channel for supermassive black hole binaries in the early universe via primordial black holes},
  author={Zhang, Saiyang and Liu, Boyuan and Bromm, Volker},
  journal={The Astrophysical Journal},
  volume={992},
  number={1},
  pages={136},
  year={2025},
  publisher={The American Astronomical Society}
}

@article{pandey2018role,
  title={Role of primordial black holes in the direct collapse scenario of supermassive black hole formation at high redshifts},
  author={Pandey, Kanhaiya L and Mangalam, A},
  eprint = "1801.06649 ",
  archivePrefix = "arXiv",
  primaryClass = "astro-ph.GA",
  journal={Journal of Astrophysics and Astronomy},
  volume={39},
  number={1},
  pages={9},
  year={2018},
  publisher={Springer}
}

@article{Johnson:2006gd,
    author = "Johnson, Jarrett L. and Bromm, Volker",
    title = "{The aftermath of the first stars: massive black holes}",
    eprint = "astro-ph/0605691",
    archivePrefix = "arXiv",
    doi = "10.1111/j.1365-2966.2006.11275.x",
    journal = "Mon. Not. Roy. Astron. Soc.",
    volume = "374",
    pages = "1557--1568",
    year = "2007"
}

@article{Dijkstra:2008jk,
    author = "Dijkstra, Mark and Haiman, Zoltan and Mesinger, Andrei and Wyithe, Stuart",
    title = "{Fluctuations in the High-Redshift Lyman-Werner Background: Close Halo Pairs as the Origin of Supermassive Black Holes}",
    eprint = "0810.0014",
    archivePrefix = "arXiv",
    primaryClass = "astro-ph",
    doi = "10.1111/j.1365-2966.2008.14031.x",
    journal = "Mon. Not. Roy. Astron. Soc.",
    volume = "391",
    pages = "1961",
    year = "2008"
}

@article{Shang:2009ij,
    author = "Shang, Cien and Bryan, Greg and Haiman, Zoltan",
    title = "{Supermassive Black Hole Formation by Direct Collapse: Keeping Protogalactic Gas H{\_}2--Free in Dark Matter Halos with Virial Temperatures T{\_}vir {\ensuremath{>}}{\textasciitilde} 10{\textasciicircum}4 K}",
    eprint = "0906.4773",
    archivePrefix = "arXiv",
    primaryClass = "astro-ph.CO",
    doi = "10.1111/j.1365-2966.2009.15960.x",
    journal = "Mon. Not. Roy. Astron. Soc.",
    volume = "402",
    pages = "1249",
    year = "2010"
}

@article{Agarwal:2012qm,
    author = "Agarwal, Bhaskar and Khochfar, Sadegh and Johnson, Jarrett L. and Neistein, Eyal and Vecchia, Claudio Dalla and Livio, Mario",
    title = "{Ubiquitous seeding of supermassive black holes by direct collapse}",
    eprint = "1205.6464",
    archivePrefix = "arXiv",
    primaryClass = "astro-ph.CO",
    doi = "10.1111/j.1365-2966.2012.21651.x",
    journal = "Mon. Not. Roy. Astron. Soc.",
    volume = "425",
    pages = "2854",
    year = "2012"
}

@article{Safranek-Shrader:2012zig,
    author = "Safranek-Shrader, Chalence and Agarwal, Meghann and Federrath, Christoph and Dubey, Anshu and Milosavljevic, Milos and Bromm, Volker",
    title = "{Star Formation in the First Galaxies I: Collapse Delayed by Lyman-Werner Radiation}",
    eprint = "1205.3835",
    archivePrefix = "arXiv",
    primaryClass = "astro-ph.CO",
    doi = "10.1111/j.1365-2966.2012.21852.x",
    journal = "Mon. Not. Roy. Astron. Soc.",
    volume = "426",
    pages = "1159",
    year = "2012"
}

@article{Regan:2014maa,
    author = "Regan, John A. and Johansson, Peter H. and Wise, John H.",
    title = "{The Direct Collapse of a Massive Black Hole Seed Under the Influence of an Anisotropic Lyman-Werner Source}",
    eprint = "1407.4472",
    archivePrefix = "arXiv",
    primaryClass = "astro-ph.GA",
    doi = "10.1088/0004-637X/795/2/137",
    journal = "Astrophys. J.",
    volume = "795",
    number = "2",
    pages = "137",
    year = "2014"
}

@article{luo2020direct,
  title={Direct collapse to supermassive black hole seeds: the critical conditions for suppression of H2 cooling},
  author={Luo, Yang and Shlosman, Isaac and Nagamine, Kentaro and Fang, Taotao},
  journal={Monthly Notices of the Royal Astronomical Society},
  eprint = "1910.07556 ",
  archivePrefix = "arXiv",
  primaryClass = "astro-ph.GA",
  volume={492},
  number={4},
  pages={4917--4926},
  year={2020},
  publisher={Oxford University Press}
}

@article{latif2022turbulent,
  title={Turbulent cold flows gave birth to the first quasars},
  journal={Nature},
  volume={607},
  number={7917},
  pages={48--51},
  year={2022},
  publisher={Nature Publishing Group UK London}
}

@article{pollack2015supermassive,
  title={Supermassive black holes from ultra-strongly self-interacting dark matter},
  author={Pollack, Jason and Spergel, David N and Steinhardt, Paul J},
  eprint = "1501.00017",
  archivePrefix = "arXiv",
  primaryClass = "astro-ph.CO",
  journal={The Astrophysical Journal},
  volume={804},
  number={2},
  pages={131},
  year={2015},
  publisher={The American Astronomical Society}
}

@article{choquette2019early,
  title={Early formation of supermassive black holes via dark matter self-interactions},
  author={Choquette, Jeremie and Cline, James M and Cornell, Jonathan M},
  eprint = "1812.05088 ",
  archivePrefix = "arXiv",
  primaryClass = "astro-ph.CO",
  journal={Journal of Cosmology and Astroparticle Physics},
  volume={2019},
  number={07},
  pages={036--036},
  year={2019}
}

@article{grant2025early,
  title={Early formation of supermassive black holes from the collapse of strongly self-interacting dark matter},
  author={Grant Roberts, M and Braff, Lila and Garg, Aarna and Profumo, Stefano and Jeltema, Tesla and O'Donnell, Jackson},
  eprint = "2410.17480 ",
  archivePrefix = "arXiv",
  primaryClass = "astro-ph.GA",
  journal={Journal of Cosmology and Astroparticle Physics},
  volume={2025},
  number={01},
  pages={060},
  year={2025},
  publisher={IOP Publishing}
}

@article{Tanaka:2013boa,
    author = "Tanaka, Takamitsu L. and Li, Miao",
    title = "{The formation of massive black holes in z {\ensuremath{\sim}} 30 dark matter haloes with large baryonic streaming velocities}",
    eprint = "1310.0859",
    archivePrefix = "arXiv",
    primaryClass = "astro-ph.CO",
    doi = "10.1093/mnras/stu042",
    journal = "Mon. Not. Roy. Astron. Soc.",
    volume = "439",
    number = "1",
    pages = "1092--1100",
    year = "2014"
}

@article{Bramberger:2015kua,
    author = "Bramberger, Sebastian F. and Brandenberger, Robert H. and Jreidini, Paul and Quintin, Jerome",
    title = "{Cosmic String Loops as the Seeds of Super-Massive Black Holes}",
    eprint = "1503.02317",
    archivePrefix = "arXiv",
    primaryClass = "astro-ph.CO",
    doi = "10.1088/1475-7516/2015/06/007",
    journal = "JCAP",
    volume = "06",
    pages = "007",
    year = "2015"
}

@article{Jiao:2025kpn,
    author = "Jiao, Hao and Brandenberger, Robert and Kamali, Vahid",
    title = "{Direct-collapse supermassive black holes from ultralight dark matter}",
    eprint = "2503.19414",
    archivePrefix = "arXiv",
    primaryClass = "astro-ph.CO",
    doi = "10.1103/9jm1-ss4t",
    journal = "Phys. Rev. D",
    volume = "112",
    number = "4",
    pages = "043546",
    year = "2025"
}

@article{Yang:2021imt,
    author = "Yang, Jinyi and others",
    title = "{Probing Early Supermassive Black Hole Growth and Quasar Evolution with Near-infrared Spectroscopy of 37 Reionization-era Quasars at 6.3 {\ensuremath{<}} z {\ensuremath{\leq}} 7.64}",
    eprint = "2109.13942",
    archivePrefix = "arXiv",
    primaryClass = "astro-ph.GA",
    doi = "10.3847/1538-4357/ac2b32",
    journal = "Astrophys. J.",
    volume = "923",
    number = "2",
    pages = "262",
    year = "2021"
}

@article{lin2026bridging,
  title={Bridging Quasars and Little Red Dots: Insights into Broad-line Active Galactic Nuclei at z= 5--8 from the First JWST COSMOS-3D Dataset},
  author={Lin, Xiaojing and Fan, Xiaohui and Wang, Feige and Sun, Fengwu and Champagne, Jaclyn B and Egami, Eiichi and Kakiichi, Koki and Lyu, Jianwei and Tee, Wei Leong and Yang, Jinyi and others},
  eprint = "2504.08039",
  archivePrefix = "arXiv",
  primaryClass = "astro-ph.GA",
  journal={The Astrophysical Journal},
  volume={996},
  number={1},
  pages={93},
  year={2026},
  publisher={The American Astronomical Society}
}

@article{Bogdan:2023ilu,
    author = "Bogdan, Akos and others",
    title = "{Evidence for heavy-seed origin of early supermassive black holes from a z{\,}{\ensuremath{\approx}}{\,}10 X-ray quasar}",
    eprint = "2305.15458",
    archivePrefix = "arXiv",
    primaryClass = "astro-ph.GA",
    doi = "10.1038/s41550-023-02111-9",
    journal = "Nature Astron.",
    volume = "8",
    number = "1",
    pages = "126--133",
    year = "2024"
}

@article{Wu:2022njo,
    author = "Wu, Jin and others",
    title = "{Demographics of z {\ensuremath{\sim}} 6 quasars in the black hole mass{\textendash}luminosity plane}",
    eprint = "2210.02518",
    archivePrefix = "arXiv",
    primaryClass = "astro-ph.GA",
    doi = "10.1093/mnras/stac2833",
    journal = "Mon. Not. Roy. Astron. Soc.",
    volume = "517",
    number = "2",
    pages = "2659--2676",
    year = "2022"
}

@article{Shen:2011ge,
    author = "Shen, Yue and Kelly, Brandon C.",
    title = "{The Demographics of Broad-Line Quasars in the Mass-Luminosity Plane. I. Testing FWHM-Based Virial Black Hole Masses}",
    eprint = "1107.4372",
    archivePrefix = "arXiv",
    primaryClass = "astro-ph.CO",
    doi = "10.1088/0004-637X/746/2/169",
    journal = "Astrophys. J.",
    volume = "746",
    pages = "169",
    year = "2012"
}

@article{he2024black,
  title={Black hole mass and Eddington-ratio distributions of less-luminous quasars at z~ 4 in the Subaru hyper suprime-cam wide field},
  author={He, Wanqiu and Akiyama, Masayuki and Enoki, Motohiro and Ichikawa, Kohei and Inayoshi, Kohei and Kashikawa, Nobunari and Kawaguchi, Toshihiro and Matsuoka, Yoshiki and Nagao, Tohru and Onoue, Masafusa and others},
  eprint = "2311.08922",
  archivePrefix = "arXiv",
  primaryClass = "astro-ph.GA",
  journal={The Astrophysical Journal},
  volume={962},
  number={2},
  pages={152},
  year={2024},
  publisher={The American Astronomical Society}
}

@article{wu2025jades,
  title={JADES-GS-z14-1: a compact, faint galaxy at z≈ 14 with weak metal lines from extremely deep JWST MIRI, NIRCam, and NIRSpec observations},
  author={Wu, Zihao and Eisenstein, Daniel J and Johnson, Benjamin D and Jakobsen, Peter and Alberts, Stacey and Arribas, Santiago and Baker, William M and Bunker, Andrew J and Carniani, Stefano and Charlot, St{\'e}phane and others},
  eprint="2507.22858",
  archivePrefix = "arXiv",
  primaryClass = "astro-ph.GA",
  journal={The Astrophysical Journal},
  volume={992},
  number={2},
  pages={212},
  year={2025},
  publisher={The American Astronomical Society}
}
\end{document}